\documentclass{article}
\usepackage{amsmath,amssymb,amsfonts}
\usepackage{graphicx}

\newtheorem{rule-of-thumb}[theorem]{Definition} 
\newcommand{\beq}{\begin{eqnarray}}
\newcommand{\eeq}{\end{eqnarray}}

\begin{document}


\title{High order eigenvalues for the Helmholtz equation in complicated non-tensor domains through Richardson Extrapolation of second order finite differences}
\author{Paolo Amore \\
Facultad de Ciencias, CUICBAS, Universidad de Colima,\\
Bernal D\'{i}az del Castillo 340, Colima, Colima, Mexico  \\
paolo.amore@gmail.com
\and
John P. Boyd \\
Department of Atmospheric, Oceanic \& Space
Science  \\University of
Michigan, 2455 Hayward Avenue, Ann Arbor MI 48109 \\
jpboyd@umich.edu; \\ http://www.engin.umich.edu:/$\sim$ jpboyd/
\and
Francisco M. Fernandez \\
INIFTA (UNLP, CCT La Plata-CONICET), \\ Divisi\'on Qu\'imica Te\'orica, Blvd. 113 y 64 (S/N), \\
Sucursal 4, Casilla de Correo 16, 1900 La Plata, Argentina \\
fernande@quimica.unlp.edu.ar
\and
Boris R\"osler \\
Facultad de Ciencias,  Universidad de Colima,\\
Bernal D\'{i}az del Castillo 340, Colima, Colima, Mexico \\
info@boris.net}




\maketitle

\begin{abstract}
We apply second order finite difference to calculate the lowest
eigenvalues of the Helmholtz equation, for complicated non-tensor
domains  in the plane, using different grids which sample exactly
the border of the domain. We show that the results obtained
applying Richardson and Pad\'e-Richardson extrapolation to a set
of finite difference eigenvalues corresponding to different grids
allows to obtain extremely precise values. When possible we have
assessed the precision of our extrapolations comparing them with
the highly precise results obtained using the method of particular
solutions. Our empirical findings suggest an asymptotic nature of
the FD series. In all the cases studied, we are able to report
numerical results which are more precise than those available in
the literature.
\end{abstract}

\section{Introduction}
\label{sec:intro}

Among the different methods for estimating the eigenvalues and eigenfunctions of the Laplacian on a finite region
of the plane, finite differences (FD) is the simplest, although the accuracy of the results obtained with this method
is limited. In particular, for domains with reentrant corners with an angle of $\pi/\alpha$, it is well known that the error of the
FD eigenvalues is dominated by a behavior  $h^{2\alpha}$ for $h \rightarrow 0$ ($h$ is the grid spacing).

The so-called L-shaped membrane [$\alpha=4/3$] is a famous example which was studied long time ago by
Fox, Henrici and Moler \cite{Fox67}.
Because of the quite slow convergence of FD in this case ($\Delta E \approx h^{4/3}$), those authors applied an alternative
method, the method of particular solutions (MPS), and, exploiting all the symmetries of the problem, they were able to
obtain the first 8 digits of the lowest eigenvalue of the L-shape correctly, $E_1 \approx 9.6397238$.
Interestingly, the paper also mentions a precise (unpublished) value  obtained by Moler and Forsythe, $E_1 \approx 9.639724$, extrapolating
the FD values obtained with very fine grids. Unfortunately, the extrapolation is neither named nor explained.

A valuable discussion of the Richardson extrapolation of FD results for the eigenvalues of the Laplacian on two dimensional
regions of the plane is contained in \cite{Kuttler84}, where it is pointed out that the correct exponents
of the asymptotic behavior of $E_1$ for $h \rightarrow 0$ must be used in the extrapolation.

The purpose of the present paper is to show that is it possible to obtain quite precise approximations to the eigenvalues
of the Laplacian on a certain class of two dimensional domains (specifically domains whose borders are sampled by the grid)
by  Richardson extrapolation of the FD results, provided that
the asymptotic behavior of the FD eigenvalues for $h \rightarrow 0$ is taken into account correctly.

The paper is organized as follows: in section \ref{sec_RE} we
provide a general discussion of Richardson extrapolation, and its
relation to the ``method of deferred corrections"; in section
\ref{Richardson}, we describe the practical implementation of the
Richardson extrapolation used in this paper; in section
\ref{Numerical} we present the numerical results obtained for
different domains, comparing them with the best results available
in the literature; finally, in section \ref{conclusions} we
summarize our findings and discuss possible directions of future
work.

\section{Richardson Extrapolation}
\label{sec_RE}

Richardson Extrapolation is interpolation of samples of a sequence $S_{n}$ by a continuous function of a continuous variable $z$ followed by
extrapolation to $z=0$ to approximate the limit of the sequence. The slowly convergent series $\sum_{n=1}^{\infty} n^{-2}$, for example, can be summed by taking the sequence of partial sums, $S_{\nu}=\sum_{n=1}^{\nu} n^{-2}$, to be samples of a function in $z \equiv 1/\nu$. In our application, the sequence is that of approximations to an eigenvalue by finite difference calculations whose asymptotic error is a series in some power of the grid spacing $h$; here $z=h^{2}$ [usually] or $z=h^{4/3}$ [for one singular application.]

The history including many independent discoveries is reviewed by Brezinski \cite{Brezinski96},
 Marchuk and Shaidurov \cite{MarchukShaidurov83}, Sidi \cite{Sidi02},
Walz \cite{Walz96} and Joyce \cite{Joyce70}.
 Christian Huyghens applied Richardson Extrapolation to estimate $\pi$ to 35 decimals from the perimeters of a sequence  of polygons with more and more sides inscribed  in the unit circle.  Richardson's (1927) paper \cite{Richardson27_I} contained a plethora of examples that was the first comprehensive display of the power of extrapolation; he claimed no novelty but credited others including an obscure Russian language paper by Bogolouboff and N. Krylov \footnote{N. Bogolouboff and N. Krylov, On the Rayleigh's principle in the theory of he differential equations of the mathematical physics and upon the Euler's method in the calculus of variations, Acad. des Sci. de l'Ukraine, Classe, Phys. Math., tonne 3, fasc. 3 (1926).} Richardson Extrapolation of eigenvalues is discussed in Pryce's book on numerical solution of Sturm-Liouville problems \cite{Pryce93}.

Richardson Extrapolation has four steps. First,  compute samples $\{ f(h_{n}) \}$ of the function being extrapolated. Second, choose a set of basis functions $\{ \phi_{j}(x) \}$ -- usually polynomials --  for an approximation
\begin{eqnarray}
f_{N}(h) \equiv \sum_{j=1}^{N} \ a_{j} \, \phi_{j}(h)
\end{eqnarray}
The coefficients $a_{j}$ can always be computed by solving a matrix problem at a cost of $O(N^{3})$ operations, and this is necessary when the $\phi_{j}$ are a mixture of polynomials and polynomials multiplied by powers of $\log(x)$, for example. However, it is faster to use Neville-Aitken interpolation to compute a  \emph{two-dimensional} array ("Richardson Table") of approximations of different $N$ formed from different subsets of the full sample set $\{ f(h_{n})  \}$. This is cheaper than matrix-solving [$O(N^{2})$ floating point operations] though this is only a small virtue because of the speed of  modern laptops. More important, extrapolation is credible only if its answers are independent of numerical choices such as $N$ and subsets of the full set of samples. More precisely, a numerical answer is believable if and only if several different values of the numerical parameters yield  the same answer to within the user chosen tolerance. The Richardson Table allows a quick search for such stable approximations. We shall return to this in analyzing each numerical example.

Various conventions are employed. A popular one is to arrange the table as a lower triangular matrix with $N$ samples of $f(z)$, the function being approximated, as the first column:
\begin{eqnarray}
R_{j,1} = f(z_{j})
\end{eqnarray}
The simple recursion is
\begin{eqnarray}
R_{j,k} =  \frac{ (z - z_{j-k-1} ) R_{j,k-1} - (z - z_{j}) R_{j-1,k-1}  }{z_{j} - z_{j-k+1}},
 \qquad k=j, (j+1), \ldots   N, \, \, j=1, 2, \ldots N
\end{eqnarray}
Each entry in column $k$ is a polynomial  of degree $(k-1)$ which interpolates
 a subset of $k$ samples. The basic step combines two polynomials that interpolate $(k-1)$ points each to generate a polynomial that interpolates at the $k$ points $\{ z_{j-k+1}, \ldots z_{j} \}$. Both generators interpolate at the $(k-2)$ points
$\{ z_{j-k+1}, \ldots z_{j} \}$, but only $R_{j,k-1}$ interpolates at $z_{j}$ while $R_{j-1,k-1}$ does not, but interpolates at $z_{j-k+1}$. It is easy to verify that

\begin{eqnarray}
R_{j,k}(z=z_{j-k+1})& = & \frac{ (z_{j-k+1} - z_{j-k-1} )  R_{j,k-1}
 - (z_{j-k+1} - z_{j})   R_{j-1,k-1}}{z_{j} - z_{j-k+1}} \\
&  = &  \frac{ - (z_{j} - z_{j})   }{z_{j} - z_{j-k+1}} R_{j-1,k-1} \\
& = &  f(z_{j-k+1} ) \qquad [\mbox{using} \, R_{j-1,k-1}(z=z_{j-k+1})= f(z_{j-k+1} )]
\end{eqnarray}

\begin{eqnarray}
R_{j,k}(z=z_{j}) &= & \frac{ (z_{j} - z_{j-k-1}  ) R_{j,k-1} - (z_{j} - z_{j})   R_{j-1,k-1}}   {z_{j} - z_{j-k+1}} \\
&= & \frac{ (z_{j} - z_{j-k-1}  }  {z_{j} - z_{j-k+1}} R_{j,k-1} \\  &  = &  f(z_{j}  \qquad [\mbox{using} \, R_{j,k-1}(z=z_{j})= f(z_{j} )]
\end{eqnarray}
\begin{eqnarray}
R_{j,k}(z=z_{j-k+1})& = & \frac{ (z_{j-k+1} - z_{j-k-1} )  R_{j,k-1}
 - (z_{j-k+1} - z_{j})   R_{j-1,k-1}}{z_{j} - z_{j-k+1}} \\
&  = &  \frac{ - (z_{j} - z_{j})   }{z_{j} - z_{j-k+1}} R_{j-1,k-1} \\
& = &  f(z_{j-k+1} ) \\
R_{j,k}(z=z_{m})& = & \frac{ (z_{m} - z_{j-k-1} ) R_{j,k-1} - (z_{m} - z_{j})  R_{j-1,k-1}}{z_{j} - z_{j-k+1}}, \qquad
m=j-k+2, \ldots {j-1} \nonumber \\
& = &   \frac{ (z_{m} - z_{j-k-1} )  - (z_{m} - z_{j}) }{z_{j} - z_{j-k+1}} f(z_{m}) \\
& = &   \frac{  - z_{j-k-1}  - z_{j}) }{z_{j} - z_{j-k+1}} f(z_{m}) \\
& = & f(z_{m})
\end{eqnarray}
where we used $R_{j,k-1}(z_{m}) = R_{j-1,k-1}(z_{m}) = f(z_{m})$ in the last lines.

For Richardson Extrapolation, we set $z=0$ and the table of polynomials becomes a lower triangular matrix of numbers.

When $z=1/n$, a reciprocal integer, Salzer gave a nice closed-form extrapolation formula in 1954 \cite{Salzer54} as well as tables
of the weights assigned to each sample in the final answer.

Sidi gives some convergence proofs in Chapter 3 of his book  \cite{Sidi02}. It is known that Richardson Extrapolation is  often exponentially
(geometrically) convergent with the error of the diagonals and  bottom rows of the table falling as $\exp(- q n)$ for some positive constant $q$
even when the power series being extrapolated is factorially divergent, as usually true when the samples are of the trapezoidal rule for different
grid spacings $h$ and the associated series in powers of $z=h^{2}$ is the Euler-Maclaurin  formula. A comprehensive theory is still lacking, however.


Richardson Extrapolation is closely related to the ``method of
deferred corrections", alternatively labelled ``correction by
higher order differences" in the (1983) book by Marchuk and
Shaidurov \cite{MarchukShaidurov83}. ``Deferred corrections" also
solves matrix problems that are the low order, usually
second-order, discretization of the problem. Deferred corrections
also promotes this low order approximation into a very high order
approximation. In contrast to Richardson Extrapolation, which
solves the low order problem repeatedly   on  a variety of
different grids, deferred corrections uses only a single grid, and
applies an iteration preconditioned by the low order
discretization \cite{Fox47,BohmerStetter84}. The residual is
evaluated by a \emph{high order} method; the accuracy of the
converged iterative solution is equally high. One grid, instead of
many, is obviously a significant advantage for deferred
correction. The method can be applied to eigenvalue problems
\cite{Chu82,ChuSpence81}.This approach has become the standard way
of generating very high order time marching schemes to pair with
spectral spatial discretizations. Dutt,  Greengard and Rokhlin
write, ``We begin by converting the original ODE into the
corresponding
  Picard equation and apply
  a deferred correction procedure in the integral
  formulation, driven by either the explicit or the implicit Euler
  marching scheme. The approach
  results in algorithms of essentially arbitrary order accuracy
for both
  non-stiff and stiff problems" \cite{DuttGreengardRokhlin00}.
Further developments of Picard integral/deferred correction time-marching can be found in  \cite{HuangJiaMinion06,LaytonMinion05,JiaHillEvansFannTaylor13}.

High order evaluation on a line in  one dimension (time) is easy, but evaluating the residual of a partial differential equation by, say, twelfth order finite differences, is a bookkeeping nightmare. The programming and debugging escalate rapidly when the domain is geometrically complicated. Furthermore, corner singularities may make higher order evaluation of the residual impossible without heroic measures \cite{Boyd99z}. For all the success of deferred correction in other applications, for eigenproblems in domains with corners Richardson Extrapolation is clearly the better way.

\section{Implementation of Richardson extrapolation}
\label{Richardson}

Suppose that we have calculated a given eigenvalue of the
Laplacian on a certain domain using finite differences for a
number of grids, which all sample the border, and with decreasing
grid spacings, $h_1 > h_2 > \dots > h_N$. Only when $h \rightarrow
0$ is the exact eigenvalue of the associated problem in the
continuum    obtained, although the eigenvalues obtained for
different (finite) grid spacing  an asymptotic behavior, which
depends on $h$; for the $k^{th}$ grid we may typically expect
\begin{eqnarray}
E_1^{(k)} = c_0 + \sum_{j=1}^\infty c_j h_k^{\alpha_j}
\label{eqRich_1}
\end{eqnarray}
where $\alpha_1 < \alpha_2 < \dots < \alpha_N$. However,
logarithms and more exotic functions have arisen       in other
problems. The exact values of these coefficients will depend on
the particular properties of the domain studied: in fact, while
integer values of $\alpha$ are associated with the discretization
of the problem ($\alpha=2,4,\dots$), rational values of $\alpha$
may also appear when reentrant corners are present (as for the
case of the L-shape where $\alpha_1=4/3$).

Using eq.~(\ref{eqRich_1}) for all grids, and with basis functions
$\phi_j$, one obtains a system of linear equations
\begin{eqnarray}
\left\{
\begin{array}{l}
E_1^{(1)} = c_0 \phi_{0} + c_1 \phi_{1}(h_{1}) + c_2 \phi_{2}(h_{1})  + \dots + c_{N-1} \phi_{N-1}(h_1) + \dots \\
E_1^{(2)} = c_0 \phi_{0} + c_1 \phi_{1}(h_{2}) c_2 \phi_{2}(h_{2}) + \dots + c_{N-1} \phi_{N-1}(h_2) + \dots \\
\dots  \\
E_1^{(N)} = c_0 \phi_{0} + c_1 \phi_{1}(h_N) + c_2 \phi_{2}(h_N)  + \dots + c_{N-1} \phi_{N-1}(h_N) + \dots \\
\end{array}\right.
\label{eqRich_2}
\end{eqnarray}
where the unknowns are the coefficients $c_j$ ($j=0,1,\dots, N-1$).

In matrix form these equations take the form
\begin{eqnarray}
\mathbf{R} \left( \begin{array}{c}
c_0 \\
c_1 \\
\dots \\
c_{N-1} \\
\end{array}\right) = \left( \begin{array}{c}
E_1^{(1)}\\
E_1^{(2)} \\
\dots \\
E_1^{(N-1)} \\
\end{array}\right)
\label{eqRich_3}
\end{eqnarray}
where
\begin{eqnarray}
\mathbf{R} \equiv  \left(
\begin{array}{ccccc}
\phi_{0} & \phi_{1}(h_1)  & \phi_{2}(h_1)  & \dots & \phi_{N-1}(h_{1})  \\
\phi_{0}  & \phi_{1}(h_2)  & \phi_{2}(h_2)   & \dots & \phi_{N-1}(h_2)  \\
\dots & \dots & \dots &  \dots & \dots \\
\phi_{0}  & \phi_{1}(h_N)  & \phi_{2}(h_N)  & \dots & \phi_{N-1}(h_N) \\
\end{array}\right)
\label{eqRich_4}
\end{eqnarray}


The solution to Eqs.~(\ref{eqRich_3}) is obtained as
\begin{eqnarray}
\left( \begin{array}{c}
c_0 \\
c_1 \\
\dots \\
c_{N-1} \\
\end{array}\right) = \mathbf{R}^{-1} \left( \begin{array}{c}
E_1^{(1)}\\
E_1^{(2)} \\
\dots \\
E_1^{(N-1)} \\
\end{array}\right)
\label{eqRich_5}
\end{eqnarray}
where the extrapolated value of $c_0$ will provide an estimate of the exact eigenvalue.

Cramer's rule can be used to obtain the coefficients $c_j$ without inverting the matrix $\mathbf{R}$; in particular
\begin{eqnarray}
c_0 &=& \frac{\left|
\begin{array}{cccc}
E_1 & \phi_{1}(h_1) & \dots & \phi_{N-1}(h_1) \\
E_2 & \phi_{1}(h_2) & \dots & \phi_{N-1}(h_2) \\
    & \dots &  & \\
E_N & \phi_{1}(h_N) & \dots & \phi_{N-1}(h_N)   \\
\end{array}
\right|}{\left|
\begin{array}{cccc}
\phi_{0} & \phi_{1}(h_1) & \dots & \phi_{N-1}(h_1) \\
\phi_{0} & \phi_{1}(h_2) & \dots & \phi_{N-1}(h_2) \\
         &  \dots        &        & \\
\phi_{0} & \phi_{1}(h_N) & \dots & \phi_{N-1}(h_N) \\
\end{array}
\right|}
\end{eqnarray}
In our numerical examples
\begin{eqnarray}
\phi_{j}(z) = z^{\alpha_{j}}
\end{eqnarray}
where $\alpha_{0}=1$ and the $\alpha_{j}$ are a monotonically  increasing sequence of positive constants.

When we apply eq.~(\ref{eqRich_1}) to the different grids, we are implicitly assuming that $\bar{h} > h_1 > \dots > h_N$, where
$\bar{h}$ is the radius of convergence of the series. However, in general $\bar{h}$ is unknown and it will only be estimated once
the first few coefficients $c_j$ have been approximated. For this reason inaccurate results could be obtained if the spacing of
one of the grids falls outside the radius of convergence of the asymptotic series. This is a common problem also of
perturbative series, which are known to be divergent in many cases.

To avoid this problem, we can  extrapolate by Pad\'{e} rational
approximation
\begin{eqnarray}
E^{(k)} = \frac{c_0 + \sum_{j=1}^N c_j h_k^{\alpha_j}}{1+ \sum_{j=1}^M d_j h_k^{\beta_j}}
\label{eqRich_6}
\end{eqnarray}

For integer exponents, $\alpha_j$ and $\beta_j$, and $N=M$, the
choice $\alpha_N = \beta_N$, would correspond to a diagonal
Pad\'e. In a general case, with $N \neq M$ and rational exponents,
we assume $\alpha_N =  \beta_M$.

Using the different grids (in this case we use $N+M+1$  grids) we obtain the system of linear equations
\beq
E^{(1)} &=& c_0 + c_1 h_1^{\alpha_1} + \dots +  c_N h_1^{\alpha_N} -  d_1 h_1^{\beta_1} E^{(1)} - \dots - d_M h_1^{\beta_M} E^{(1)} \nonumber \\
E^{(2)} &=& c_0 + c_1 h_2^{\alpha_1} + \dots +  c_N h_2^{\alpha_N} -  d_1 h_2^{\beta_1} E^{(2)} - \dots - d_M h_2^{\beta_M} E^{(2)} \nonumber \\
\dots  &=& \dots \nonumber \\
E^{(N+M+1)} &=& c_0 + c_1 h_{N+M+1}^{\alpha_1} + \dots +  c_N h_{N+M+1}^{\alpha_N} -  d_1 h_{N+M+1}^{\beta_1} E^{(N+M+1)} \nonumber \\
&-& \dots - d_M h_{N+M+1}^{\beta_M} E^{(N+M+1)} \nonumber
\eeq
which can be cast in matrix form as
\begin{eqnarray}
\tilde{\mathbf{R}} \left( \begin{array}{c}
c_0 \\
c_1 \\
\dots \\
c_{N} \\
d_1 \\
\dots \\
d_M \\
\end{array}\right) = \left( \begin{array}{c}
E_1^{(1)}\\
E_1^{(2)} \\
\dots \\
E_1^{(M+N+1)} \\
\end{array}\right)
\label{eqRich_7}
\end{eqnarray}
where
\begin{eqnarray}
\tilde{\mathbf{R}} \equiv  \left(
\begin{array}{cccccccc}
1 & h_1^{\alpha_1} & \dots & h_1^{\alpha_N} &  - h_1^{\beta_1} E^{(1)}  &   \dots & - h_1^{\beta_M} E^{(1)} \\
1 & h_1^{\alpha_1} & \dots & h_2^{\alpha_N} &  - h_2^{\beta_1} E^{(2)}  &   \dots & - h_2^{\beta_M} E^{(2)} \\
\dots & \dots  &  \dots & \dots &  \dots & \dots & \dots \\
1 & h_{N+M+1}^{\alpha_1}  & \dots & h_{N+M+1}^{\alpha_N} &  - h_{N+M+1}^{\beta_1} E^{(N+M+1)}  &   \dots & - h_{N+M+1}^{\beta_M} E^{(N+M+1)} \\
\end{array}\right)
\label{eqRich_8}
\end{eqnarray}

The solutions to these equations are found inverting $\tilde{R}$
\beq
\left( \begin{array}{c}
c_0 \\
c_1 \\
\dots \\
c_{N} \\
d_1 \\
\dots \\
d_M \\
\end{array}\right) &=& \tilde{\mathbf{R}}^{-1} \left( \begin{array}{c}
E_1^{(1)}\\
E_1^{(2)} \\
\dots \\
E_1^{(M+N+1)} \\
\end{array}\right)
\label{eqRich_9}
\eeq
or using Cramer's rule once again.

\section{Numerical results}
\label{Numerical}

To apply the extrapolation schemes described in the previous section we need to calculate accurately the FD eigenvalues for a
series of grids. We consider different domains, with borders which can be sampled by a square grid and with different reentrant angles.

\subsection{L-shaped domain}

\begin{figure}[!t]
\begin{center}
\includegraphics[width=4cm]{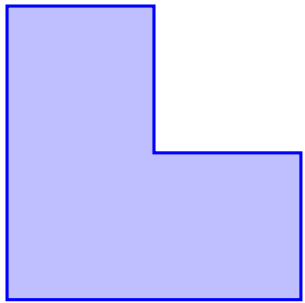}
\caption{L-shaped region}
\label{Fig_L}
\end{center}
\end{figure}

We consider the L-shaped region $\Omega \equiv \left\{ |x| < 1, |y|<1 \right\} - \left\{ 0 \leq x < 1, 0 \leq y<1 \right\}$,
represented in Fig.~\ref{Fig_L}.
Using finite differences and a five-points approximation to the Laplacian, the Helmholtz equation on $\Omega$ is solved with
Dirichlet boundary conditions on $\partial\Omega$ for a series of grids with an increasing number of points.  We have exploited the symmetry
of the domain, to obtain separately the even and odd modes of the L-shape.

Our numerical calculations consist of two sets:
\begin{itemize}
\item A calculation of the lowest eigenvalue of the L, using $124$ grids with spacing $h=1/N_0$ and $N_0=10, \dots, 133$.
The finite difference results of this set are obtained using the "Conjugate Gradient Method" (CGM), as described
in Ref.~\cite{Nightingale93},  and they are accurate to $220$ digits;

\item A calculation of the lowest $100$ eigenvalues of the L, using 100 grids with spacing $h=1/N_0$ and $N_0=10, \dots, 109$.
The finite difference results of this set are obtained using the internal Mathematica command \verb Eigenvalues   and
they are accurate to $60$ digits.
\end{itemize}

In Table \ref{tab_results_L} we report the available estimates of the lowest eigenvalue of the L-shape in the literature,
including the results of the present work.

\begin{table}[t]
\caption{Available estimates of the lowest eigenvalue of the L-shape (smaller fonts are used for the last three values, to allow fitting
the results in the column).}
\bigskip
\label{tab_results_L}
\begin{center}
\begin{tabular}{|l|l|}
\hline
 & $E_1$ \\
\hline
Reid and Walsh \cite{Reid65}                 &  9.63972\\
Fox, Henrici and Moler \cite{Fox67}          &  9.6397238 \\
Mason \cite{Mason67}                         &  9.6397\\
Sideridis \cite{Sideridis84}                 &  9.6395 \\
Schiff \cite{Schiff88}                       &  9.659 \\
Christiansen and Petersen \cite{ChristiansenPetersen89} & 9.6397238{\tiny 3991} \\
Still \cite{Still03}                         &  $9 639723^{96}_{71}$ \\
Betcke and Trefethen \cite{Betcke05}         &  9.6397238440219 \\
Amore \cite{Amore08}                         &  9.6397238440 \\
Yuan and He \cite{He09}                      &  $9.63972384^{44}_{04}$ \\
this work {\bf\tiny (Richardson)}                            &  {\tiny  9.63972384402194105271145926236482315626728952582190645} \\
this work {\bf\tiny (Pad\'e-Richardson)}                  &  {\tiny  9.6397238440219410527114592623648231562672895258219064561095797005640} \\
this work {\bf\tiny (MPS)}                   &  {\tiny  9.639723844021941052711459262364823156267289525821906456109579700564036} \\
\hline
\end{tabular}
\end{center}
\bigskip\bigskip
\end{table}

As we have mentioned before, the convergence of the numerical results is affected by the presence
of a reentrant corner and the finite-difference eigenvalue $E(h)$ behaves for $h\rightarrow 0$
as\cite{Donnelly69,Kuttler84}
\beq
E(h) = E(0) + a h^{4/3} + \dots
\eeq
where $E(0)$  is the  eigenvalue of the Laplacian in the continuum. For the related problem of
a H-shaped membrane, Donnelly~\cite{Donnelly69}  conjectured  the asymptotic behavior
\beq
E(h) = E(0) + a h^{4/3} + b h^2 + c h^{10/3} + d h^4 + \dots
\eeq
for the fundamental eigenvalue~\footnote{Since the H-shaped domain contains the same reentrant
angle of the L-shape, we assume the same asymptotic law for both domains.}. This behavior was also
used by Christiansen and Petersen \cite{ChristiansenPetersen89} to perform a Richardson extrapolation
of the finite difference results for the L-shape (see Table \ref{tab_results_L}).

The results obtained extrapolating the FD sequences can be
compared with the precise results obtained with the "method of
multiple solutions" (MPS)\cite{Fox67}. Table \ref{tab_MPS} reports
the first $25$ eigenvalues of the L-shape obtained with the MPS
(for the case of the first eigenvalue we have used 545 points
evenly spaced, which allow one to obtain 70 digits of precision,
for the remaining cases we have used $425$ points, which allows an
accuracy of about 50 digits). The eigenvalues marked with
$\dagger$ are known exactly and correspond to modes of a square.
The MPS  has been implemented in Mathematica 10 \cite{math15},
taking advantage of Mathematica's ability to work with arbitrary
precision numbers or with a large number of digits (in our case
typically numbers are specified to 100 digits).

We will use these values to establish the accuracy of the approximate values of $E_n$ obtained by applying
four different extrapolation schemes, differing in the choice of the exponents:
\begin{itemize}
\item Extrapolation ${\rm i}$
\beq
E(h) &=& E(0)  + \sum_{n=1}^\infty c^{(i)}_n h^{2n}
\approx E(0)  + c^{(i)}_1 h^{2} + c^{(i)}_2 h^{4}   + O(h^6)
\eeq

\item Extrapolation ${\rm ii}$
\beq
E(h) &=& E(0) + \sum_{n=1}^\infty c^{(ii)}_n h^{n}
\approx  E(0)  + c^{(ii)}_1 h^{2} + c^{(ii)}_2 h^{3}   + O(h^4)
\eeq

\item Extrapolation ${\rm iii}$ (Donnelly, Ref.~\cite{Donnelly69})
\beq
E(h) &=& E(0)  + \sum_{n=1}^\infty \left[ c^{(iii)}_{2n-1} h^{2n-2/3} +  c^{(iii)}_{2n} h^{2n} \right] \nonumber \\
&\approx& E(0)  + c^{(iii)}_1 h^{4/3} + c^{(iii)}_2 h^{2} + c^{(iii)}_3 h^{10/3}  + O(h^{4})
\eeq

\item Extrapolation ${\rm iv}$
\beq
E(h) &=& E(0)  + \sum_{n=1}^\infty c^{(iv)}_n h^{2(n+1)/3} \nonumber \\
&\approx& E(0)  + c^{(iv)}_1 h^{4/3} + c^{(iv)}_2 h^{2} + c^{(iv)}_3 h^{8/3}  + O(h^{10/3})
\eeq
\end{itemize}

The first two schemes only use integer exponents and are expected to be accurate only for the modes of the
L-shape which are also modes of the square.

\begin{figure}
\begin{center}
\includegraphics[width=7cm]{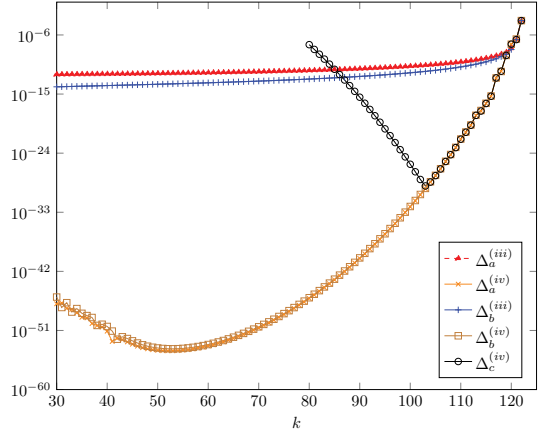}
\caption{Error over the first eigenvalue of the L-shaped region. The first two curves report the difference between the values obtained with Richardson extrapolation of $124-k$ grids, respectively using scheme ${\rm iii}$ and ${\rm iv}$, and the precise value that
we have obtained with the MPS; the last two curves report the difference between the values obtained with Richardson extrapolation of $124-k$ grids and the values obtained with Richardson extrapolation of $124-k-1$ grids, respectively using scheme ${\rm iii}$ and ${\rm iv}$. This difference essentially provides the number of stable digits achieved. In the first four curves the FD eigenvalues are obtained with
an accuracy of 220 digits; the last curve is analogous to the second one, limiting the accuracy of the FD eigenvalues to 60 digits}
\label{Fig_L_error}
\end{center}
\end{figure}

Figure \ref{Fig_L_error} displays the error $|E_1^{(extra)} -E_1^{(MPS)}|$ for the
lowest eigenvalue of the L-shaped region, using the third and fourth extrapolation schemes.
Here
\begin{eqnarray}
\Delta_a &=& |\mathcal{R}^{(k,124)} (E_1) - E_1^{(MPS)}| \\
\Delta_b &=& |\mathcal{R}^{(k,124)} (E_1) - \mathcal{R}^{(k-1,124)} (E_1)|
\end{eqnarray}
where the superscripts $(iii)$ and $(iv)$ refer to the series used and the FD eigenvalues are accurate to 220
digits. The values $\Delta_c^{(iv)}$ are the analogous of $\Delta_a^{(iv)}$, but using FD eigenvalues are accurate to 60 digits.

The approximations obtained with the first two schemes, which do
not use rational exponents, are very poor for this mode.

In particular, the extrapolated values in the four cases are
\begin{itemize}
\item Extrapolation ${\rm i}$
\begin{equation}
E_1 \approx \underline{9.639}8
\end{equation}

\item Extrapolation ${\rm ii}$
\begin{equation}
E_1 \approx \underline{9.6397}327
\end{equation}

\item Extrapolation ${\rm iiii}$
\begin{equation}
E_1 \approx \underline{9.639723844021}1929465
\end{equation}

\item Extrapolation ${\rm iv}$ (corresponding to the minimum in Fig.~\ref{Fig_L_error})
\begin{equation}
E_1 \approx \underline{9.639723844021941052711459262364823156267289525821906456}458
\end{equation}
\end{itemize}

Remarkably, the fourth scheme provides the first $55$ digits of $E_1$ for the L-shape correctly,
suggesting that the the exact asymptotic behavior of the finite difference eigenvalues, for $h \rightarrow 0$,
is $E(h) = E(0) + \sum_{n=1}^\infty c^{(iv)}_n h^{2(n+1)/3}$.

In correspondence to the minimum of Fig.~\ref{Fig_L_error} we have calculated the first few coefficients of the asymptotic series
for the eigenvalue of the fundamental mode; the expansion reads (underlined digits are expected to have converged)
\beq
E(h)  &\approx& \underline{9.63972384402194105271145926236482315626728952582190645}6 \nonumber \\
&+&  \underline{2.19759909080385142157537952672409583683648557094}5 \ h^{4/3} \nonumber \\
&-& \underline{5.254349649878412271190008297029240841285038851}0  \ h^{2} \nonumber \\
&-& \underline{0.045716100985365949827658978449794728350032}8  \ h^{8/3}  \nonumber \\
&-& \underline{1.946468144036811059220897747699440650587}7 \ h^{10/3} \nonumber \\
&+& \underline{1.125074754927755172836371946813777186}1 \ h^4 \nonumber \\
&-& \underline{0.2147544087374345021476728527871998}5 \ h^{14/3}  \nonumber \\
&+& \underline{0.35588422353456505262712958896229}4 \ h^{16/3} \nonumber \\
&+& \underline{0.006403070910486707732478038349}7 \  h^6 \nonumber \\
&+& \underline{0.038286091425541761563564936}0 \ h^{20/3} \nonumber \\
&-& \underline{0.0730523282127573068239088}6 \ h^{22/3} + \dots
\label{asym}
\eeq

The behavior of the error in Fig.~\ref{Fig_L_error} suggests that the FD series is asymptotic.
Therefore, if one picks a set of grids with spacings $h_1> h_2 > \dots$, it is convenient to perform
an extrapolation  using the grids up to a given spacing $h_N$ where the error reaches a minimum.

\begin{figure}
\begin{center}
\includegraphics[width=7cm]{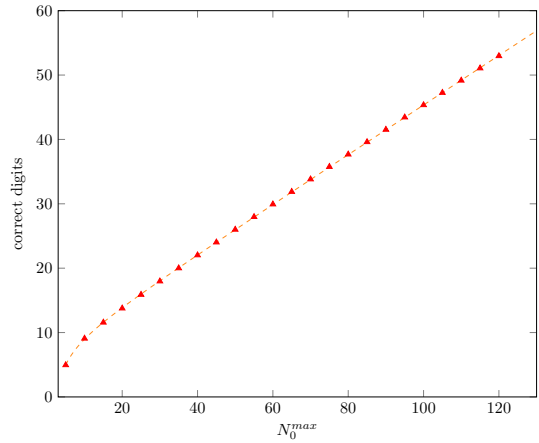}
\caption{Correct digits of the lowest eigenvalue of the  L-shaped membrane  obtained with Richardson extrapolation using a set of FD grids with a
smallest spacing $h_{min} = 1/N_0^{max}$. Notice that the number of grids used for a given $h_{min}$ depends on $h_{min}$ itself
(see Fig.~\ref{Fig_2b}). The dashed curve is the fit $f(n) = 7.23166\, +0.383229 n-\frac{20.9176}{n}$. The FD eigenvalues used
in the extrapolation were computing using 220 decimal digit floating point arithmetic.}
\label{Fig_2}
\end{center}
\end{figure}

\begin{figure}
\begin{center}
\includegraphics[width=7cm]{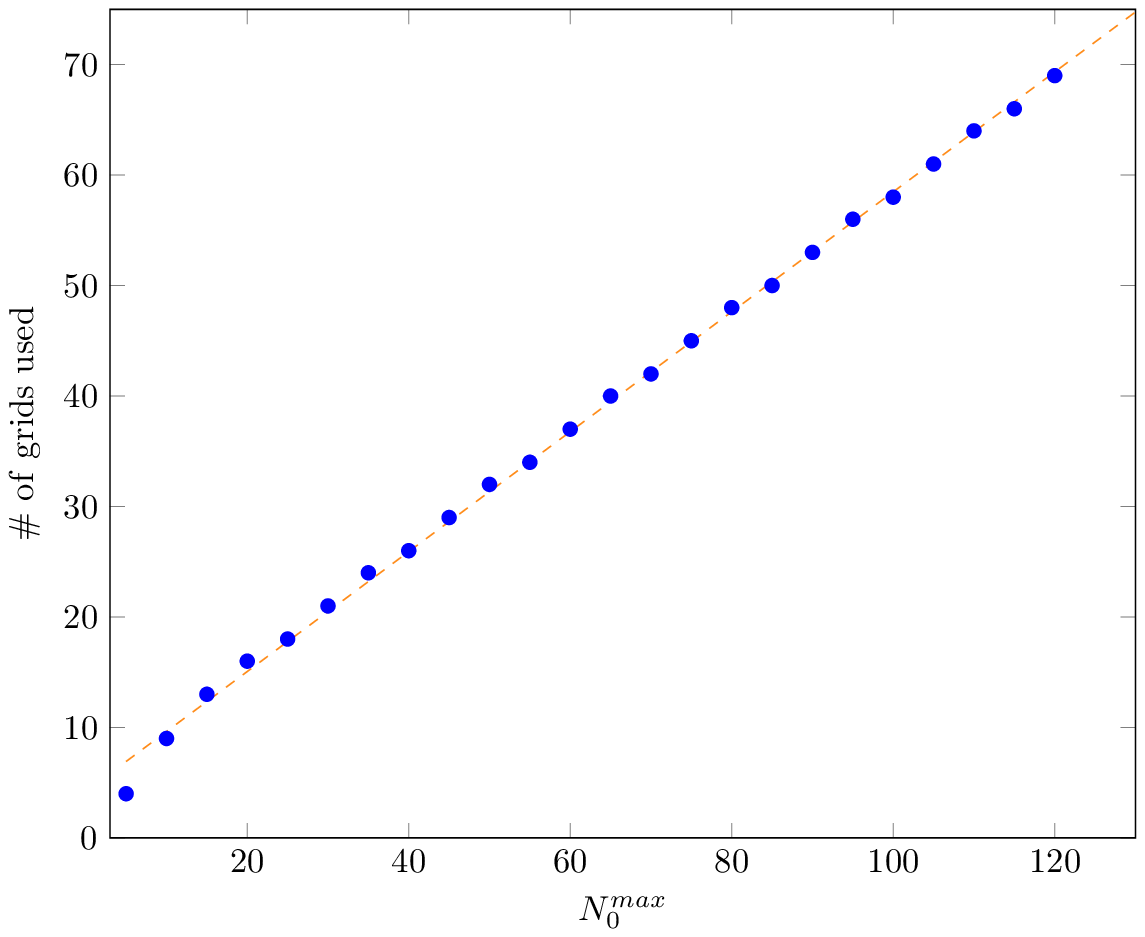}
\caption{Optimal number of FD grids used for a set  of FD grids with smallest spacing
$h_{min} = 1/N_0^{max}$. The dashed curve is the fit $g(n) =0.542696 n+4.20652$. The FD eigenvalues used in the extrapolation were computed in multiple precision floating point arithmetic with a    precision of
220 decimal digits.}
\label{Fig_2b}
\end{center}
\end{figure}

This behavior, however, does not limit the number of accurate digits of the eigenvalue that one can obtain using
the Richardson extrapolation. This is illustrated in Figs.~\ref{Fig_2} and \ref{Fig_2b}: the first figure
is obtained extrapolating the FD results of a set with smallest spacing $h_{min}$ and determining the minimum error
over the extrapolated eigenvalue (which will correspond to the minimum observed in Fig.~\ref{Fig_L_error}). In this case
we observe that the number of accurate digits of the extrapolated eigenvalue grows linearly for $N_0 \gg 1$. Of course
this behavior will be lost when the number of digits of the FD eigenvalue is not sufficient (see for example, the last
curve of Fig.~\ref{Fig_L_error}, where the FD eigenvalue are only accurate to 60 digits).
Fig.~\ref{Fig_2b} illustrates the fact that, as $h_{min}$ gets smaller and smaller, the number of grids used in the
optimal extrapolation also grows linearly.

\begin{figure}
\begin{center}
\includegraphics[width=7cm]{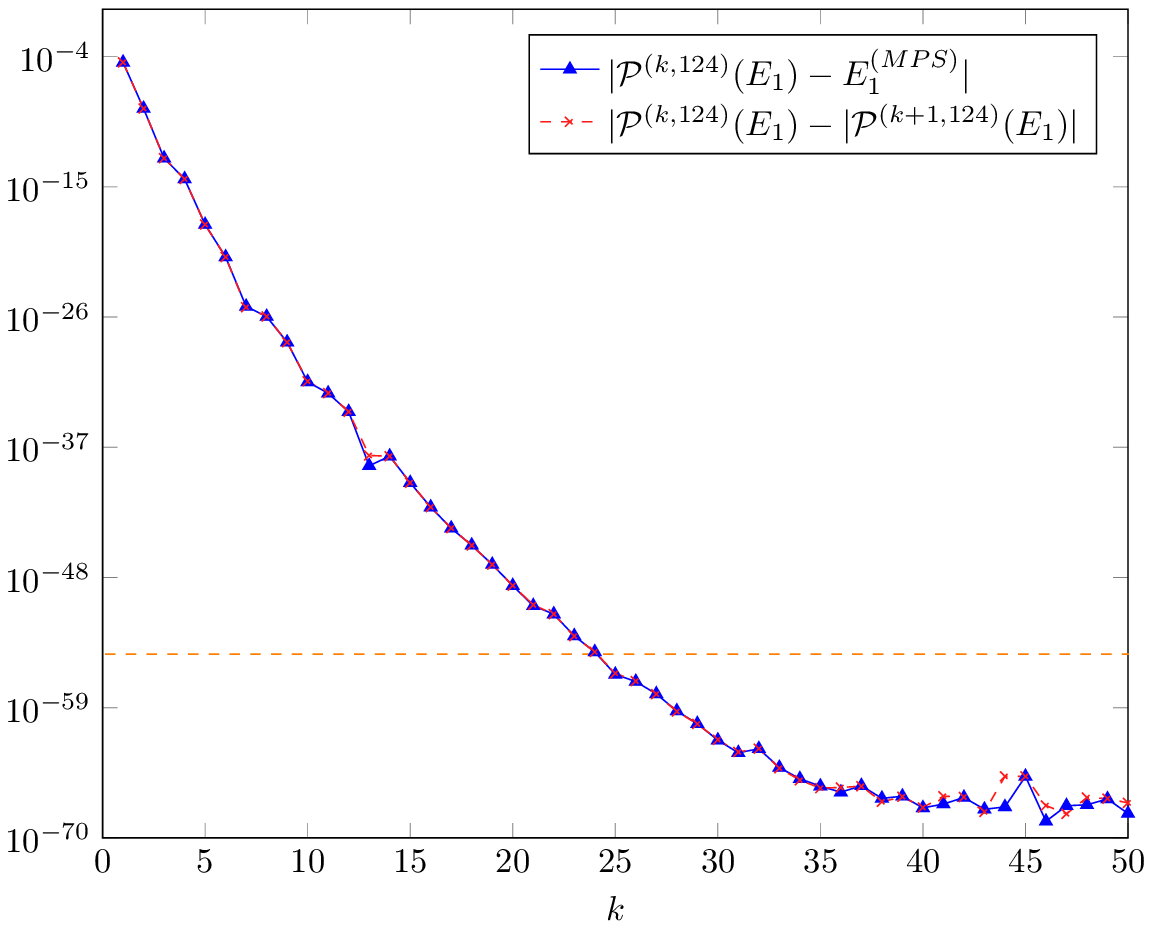}
\caption{Error in the first eigenvalue of the L-shpaed domain using the diagonal Pad\'{e}-Richardson Extrapolation  $\mathcal{P}^{(k,124)}$.
The horizontal line corresponds to the minimal error obtained with the Richardson extrapolation, corresponding
to the minimum in Fig.~\ref{Fig_L_error}.}
\label{Fig_pade}
\end{center}
\end{figure}

In Fig.~\ref{Fig_pade} we have applied the Pad\'e-Richardson
extrapolation to calculate the error over the fundamental
eigenvalue of the L. Here   $\mathcal{P}^{(k,124)}$ indicates the
diagonal Pad\'e with $2k+1$ coefficients, which uses the grids
going from $124-2k$ to $124$. The horizontal line corresponds to
the lowest error obtained with the Richardson extrapolation, i.e.
to the minimum of Fig.~\ref{Fig_L_error}. The errors are obtained
using as a reference the precise estimate obtained using the MPS
with 545 points distributed on the border, which is expected to
have at least 70 correct digits (see Table \ref{tab_results_L}).

The result obtained with the Pad\'e-Richardson extrapolation
contains 13 extra digits of accuracy with respect to the result
obtained with the Richardson extrapolation alone!!

The same analysis can be carried out for the eigenvalue of the first excited mode of the L-shaped membrane, which is odd with respect to reflection
about the line $y=x$; also in this case, the fourth scheme is the appropriate one and  the asymptotic expansion is obtained
\beq
E(h)  &\approx& \underline{15.19725192645434327487838213300054590}06  +  3.18 \cdot  10^{-32} \ h^{4/3} \nonumber \\
&-& \underline{12.565568615260003775714180770}53 \ h^{2}  \nonumber \\
&-& \underline{2.2529040988480935561491817}46 \ h^{8/3}  -9.9 \cdot 10^{-25} \ h^{10/3} \nonumber \\
&+& \underline{3.932508901213713526500}07 \ h^4 + \underline{1.1289729308101123792}42 \ h^{14/3}  \nonumber \\
&+& \underline{0.950164117523872693}21 \ h^{16/3} - \underline{1.36902891120799}30 \  h^6 \nonumber \\
&-& \underline{0.0740361191169}66 \ h^{20/3} + \underline{0.772459685647}50 \ h^{22/3} + \dots
\label{asym2}
\eeq

Notice that in this case we have used the less precise set of FD values, which were computed only in 60 digit floating point arithmetic: the eigenvalue of the
first excited state is now reproduced with ``just" 37 correct digits.

This result clearly shows that the coefficients of the terms $h^{4/3}$ and $h^{10/3}$ must vanish: in particular it is easy to
understand the absence of $h^{4/3}$ since the mode that we are calculating is the fundamental eigenmode of the
desymmetrized region obeying Dirichlet boundary conditions on $y=x$. In this case the reentrant corner is $\pi/\alpha = 3\pi/4$ and therefore
$2\alpha=8/3$.

With this simple observation, eliminating $4/3$ and $10/3$ from the exponents used in the extrapolation scheme,
we are able to obtain 3 more digits of $E_2$
\begin{equation}
\log_{10} \frac{1}{|E_2^{({\rm RE})}-E_2^{\rm (MPS)}|} = 40.8 \nonumber
\end{equation}

Even more digits can be obtained using the Pad\'{e}-Richardson scheme, without the exponents $4/3$ and $10/3$: in this case
\begin{equation}
\log_{10} \frac{1}{|E_2^{({\rm PRE})}-E_2^{\rm (MPS)}|} = 45.8 \nonumber
\end{equation}

\begin{table}[!htbp]
\caption{Lowest 25 eigenvalues of the L-shaped domain obtained with the MPS using $425$ points evenly spaced on the border. The eigenvalues
marked with $\dagger$ are known exactly; the first eigenvalue, marked with $*$, has been obtained using the MPS with $545$ points.}
\bigskip
\label{tab_MPS}
\begin{center}
\begin{tabular}{|c|l|}
\hline
$n$ & $E_n^{({\rm MPS})}$ \\
\hline
1$^*$          &  9.639723844021941052711459262364823156267289525821906456109579700564036 \\
2           &  15.197251926454343274878382133000545900777179939609   \\
3$^\dagger$ &  $2 \pi^2$ \\
4           &  29.521481114144883298220387998949268230835182037083   \\
5           &  31.912635957137762200327505645485619891180683442197   \\
6           &  41.474509890214922338810104064796906887679915692804   \\
7           &  44.948487781351230152829670239630032397049780134665   \\
8$^\dagger$ &  $5 \pi^2$ \\
9$^\dagger$ &  $5 \pi^2$ \\
10          &  56.709609887385120714216741638492259079610565870838   \\
11          &  65.376535709845878509384400627738811907191161706097   \\
12          &  71.057755648513529930798223378765313509589316160842   \\
13          &  71.572679680336556014706999077329408038228565031443   \\
14          &  $8 \pi^2$ \\
15          &  89.301668351960185629207557215836143584908527108716   \\
16          &  92.306906763049247832266397297040944898714305036279   \\
17          &  97.380722646021860253461536778106579066564981169123   \\
18          &  $10 \pi^2$ \\
19          &  $10 \pi^2$ \\
20          &  101.60529408377871548543481415097538087072356189211    \\
21          &  112.36860922562569413546584663077376004912074741174    \\
22          &  115.52017309466770886932756039014897616475657545671    \\
23          &  $13 \pi^2$ \\
24          &  $13 \pi^2$ \\
25          &  130.11902885096790256577606801292831058988583848246    \\
\hline
\end{tabular}
\end{center}
\bigskip\bigskip
\end{table}

~\begin{table}[!htbp] \caption{Correct digits of the first 25
eigenvalues of the L-shaped domain, obtained applying the
Richardson and Richardson-Pad\'e extrapolations to FD eigenvalues.
The values marked with the $\dagger$ correspond to  eigenstates of
the square. The first eigenvalue has been obtained extrapolating
the FD eigenvalues of $124$ grids, obtained with a floating point
precision of 220 digits. }
\bigskip
\label{table3}
\begin{center}
\begin{tabular}{|l|c|c|c|c|}
\hline
$n$ & scheme & $\log_{10} \frac{1}{|E_n^{({\rm RE})}-E_n^{\rm (MPS)}|}$ & $\log_{10} \frac{1}{|E_n^{({\rm PRE})}-E_n^{\rm (MPS)}|}$ & parity \\
\hline
1$^*$           & ${\rm iv}$  & 54.5 & 67.5  & ${\rm even}$  \\
2           & ${\rm iv}$  & 40.8  & 45.8  & ${\rm odd}$ \\
3$^\dagger$ & ${\rm i}$   & 62.9  & 73.1  & ${\rm even}$  \\
4           & ${\rm iv}$  & 37.1  & 45.8  & ${\rm odd}$ \\
5           & ${\rm iv}$  & 35.9  & 42.6  & ${\rm even}$ \\
6           & ${\rm iv}$  & 35.1 & 42.2  & ${\rm even}$ \\
7           & ${\rm iv}$  & 36.7 & 44.5  & ${\rm odd}$ \\
8$^\dagger$ & ${\rm i}$   & 60.6 & 73.9  & ${\rm odd}$ \\
9$^\dagger$ & ${\rm i}$   & 60.8 & 73.8  & ${\rm even}$ \\
10          & ${\rm iv}$  & 35.2 & 41.9  & ${\rm even}$  \\
11          & ${\rm iv}$  & 34.5 & 42.6  & ${\rm odd}$  \\
12          & ${\rm iv}$  & 34.8 & 42.2  & ${\rm even}$ \\
13          & ${\rm iv}$  & 34.4 & 42.6  & ${\rm odd}$ \\
14$^\dagger$ & ${\rm i}$  & 60.3 & 73.2  & ${\rm even}$  \\
15           & ${\rm iv}$ & 33.4 & 41.3  & ${\rm even}$ \\
16           & ${\rm iv}$ & 30.8 & 39.9  & ${\rm odd}$ \\
17           & ${\rm iv}$ & 30.6 & 39.3  & ${\rm odd}$ \\
18$^\dagger$ & ${\rm i}$  & 60.3 & 74.0  & ${\rm odd}$  \\
19$^\dagger$ & ${\rm i}$  & 59.5 & 73.9  & ${\rm even}$ \\
20          & ${\rm iv}$ & 33.0 & 40.7  & ${\rm even}$ \\
21           & ${\rm iv}$ & 32.6 & 40.0  & ${\rm even}$ \\
22           & ${\rm iv}$ & 33.7 & 42.6  & ${\rm odd}$ \\
23$^\dagger$ & ${\rm i}$  & 59.6 & 73.6  & ${\rm odd}$  \\
24$^\dagger$ & ${\rm i}$  & 59.7 & 73.2  & ${\rm even}$  \\
25           & ${\rm iv}$ & 33.3 & 43.4  & ${\rm odd}$  \\
\hline
\end{tabular}
\end{center}
\bigskip\bigskip
\end{table}

\subsection{H-shaped domain}

We now consider a domain with the shape of H, displayed in Fig.~\ref{Fig_2}, originally
studied by Donnelly \cite{Donnelly69} using the method of particular solutions (MPS) and
finite differences (FD). As we have already mentioned in the previous section,
the author conjectured that the FD eigenvalues, corresponding to a given grid spacing $h$,
behave as
\beq
E(h) = E(0) + a h^{4/3} + b h^2 + c h^{10/3} + d h^4 + \dots
\eeq
where $E(0)$ is the corresponding eigenvalue of the Laplacian in the continuum and the exponent
$4/3$ is determined by the presence of a reentrant corner $3\pi/2$~\cite{Donnelly69,Kuttler84}.

As for the L-shape, we want to obtain a precise estimate of the lowest eigenvalues for
this problem, using a sequence of FD eigenvalues, obtained for different grids. Notice that
the eigenfunctions of the Laplacian on this domain can be classified according to four different symmetry classes,
even-even, even-odd, odd-even and odd-odd with respect to reflection about the $x$ and $y$ axes.
By working separately on the modes belonging to each class, the computational complexity of the problem can
be reduced and finer grids can be studied. Our present analysis, in particular, is limited to the even-even modes.
The spacing of the grid is chosen so that the border of the H-shaped is sampled exactly and it corresponds to
$h_k = 3/2/(9+3 (k-1))$, with $k=1,2,\dots$. We have calculated the first $25$ eigenvalues of the even-even modes of the
H-shape with a floating point  precision of $60$ digits, for the grids corresponding to $k=1, 2, \dots, 40$.

Our results for the lowest eigenvalue should be compared with those of Donnelly~\cite{Donnelly69}
\begin{equation}
E_1^{(Donnelly)} = 7.7330889
\end{equation}
and, more recently, of Betcke and Trefethen~\cite{Betcke05}
\begin{equation}
E_1^{(BT)} = 7.7330888559
\end{equation}

\begin{figure}
\begin{center}
\includegraphics[width=4cm]{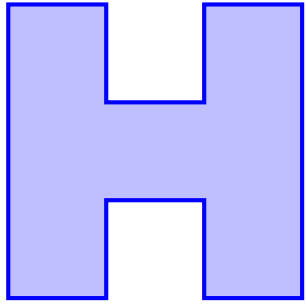}
\caption{H-shaped region}
\label{Fig_H}
\end{center}
\end{figure}

\begin{figure}
\begin{center}
\includegraphics[width=8cm]{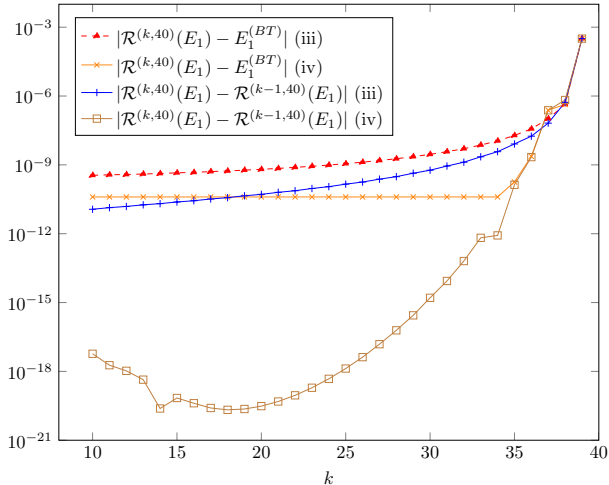}
\caption{Error over the first eigenvalue of the H-shaped region. The first two curves report the difference between the values obtained with Richardson extrapolation of $40-k$ grids, respectively using scheme ${\rm iii}$ and ${\rm iv}$, and the precise value of Betcke and Trefethen \cite{Betcke05}; the last two curves report the difference between the values obtained with Richardson extrapolation of $40-k$ grids and the values obtained with Richardson extrapolation of $40-k-1$ grids, respectively using scheme ${\rm iii}$ and ${\rm iv}$. This difference essentially provides the number of stable digits
achieved. }
\label{Fig_H_error}
\end{center}
\end{figure}

In Fig.~\ref{Fig_H_error} we report the error over the first eigenvalue of the H-shape. The first two curves report the difference between the values obtained with Richardson extrapolation of $40-k$ grids, respectively using scheme ${\rm iii}$ and ${\rm iv}$, and the precise value of Betcke and Trefethen \cite{Betcke05}. However, since the results of Ref.~\cite{Betcke05} are not sufficiently precise, it is convenient to estimate the error
using the difference between the values obtained with Richardson extrapolation of $40-k$ grids and the values obtained with Richardson extrapolation of $40-k-1$ grids, respectively using scheme ${\rm iii}$ and ${\rm iv}$. This difference essentially provides the number of stable digits
achieved.  Notice that the second curve rapidly reaches a plateau, for $k \leq 34$, signaling that in this range the extrapolated results
are more precise than those of Ref.~\cite{Betcke05}.

The figure clearly shows that the asymptotic behavior conjectured by Donnelly in Ref.~\cite{Donnelly69} is not correct;
our best estimate of the fundamental eigenvalue corresponds to the last curve in Fig.~\ref{Fig_H_error} (i.e. scheme {\rm iv}) for $k=18$:
\begin{equation}
E_1 = 7.7330888559426190667
\end{equation}
where all the digits are believed to be correct.

In table ~\ref{tab_H} we report the approximate values of the first 24 eigenvalues of the even-even modes of the H-shape obtained using
Richardson extrapolation. It is particularly interesting to consider the value for the mode 24, which has the lowest precision.
The coefficients of the asymptotic series obtained from the Richardson extrapolation are (underlined digits are expected to have converged)
\begin{eqnarray}
E(h) &\approx& \underline{194.7347257248}53 + \underline{1.2880}50 h^{4/3} - \underline{2861.993}46 h^2 \nonumber \\
&-& \underline{25.7}61 h^{8/3} + \underline{51}5.6 h^{10/3} + \underline{14}691.3 h^4  + \dots
\end{eqnarray}

The coefficients of this series, although determined with less precision than in the cases discussed earlier for the L-shape,
clearly suggest the presence of a smaller radius of convergence, which drastically affects the accuracy of the calculation.

\begin{table}[!htbp]
\caption{Lowest 24 eigenvalues of even-even modes of the H-shaped domain obtained using Richardson extrapolation with set ${\rm iv}$
(the sets marked with $\dagger$ are eigenstates of the square and are extrapolated using set ${\rm i}$).}
\bigskip
\label{tab_H}
\begin{center}
\begin{tabular}{|c|l|}
\hline
$n$ & $E_n^{({\rm Richardson})}$ \\
\hline
1            &  7.7330888559426190667 \\
2            &  14.30522996107150163018552 \\
3$^\dagger$  &  19.73920880217871723766898199975230227062739 \\
4            &  33.0048892952083545188 \\
5            &  37.2054234400574157525 \\
6            &  46.2961910861973723751 \\
7            &  58.7501048292892847997 \\
8            &  63.113298546574958190  \\
9            &  67.43457224647486521   \\
10           &  85.80372978847046992   \\
11           &  92.12485042399187898   \\
12           &  95.7615825533281487    \\
13$^\dagger$ &  98.696044010893586188344909998 \\
14           &  112.42755013401679304  \\
15           &  122.557976404091254965 \\
16           &  133.5364354179283      \\
17           &  139.4282184592822      \\
18           &  142.4312241050896      \\
19           &  150.543062476658690    \\
20           &  164.339040164448839    \\
21           &  171.85972578742946     \\
22$^\dagger$ &  177.652879219608455139020837997770 \\
23           &  180.46602205029118     \\
24           &  194.7347257248         \\
\hline
\end{tabular}
\end{center}
\bigskip\bigskip
\end{table}


\subsection{Isospectral domains}

\begin{figure}
\begin{center}
\includegraphics[width=4cm]{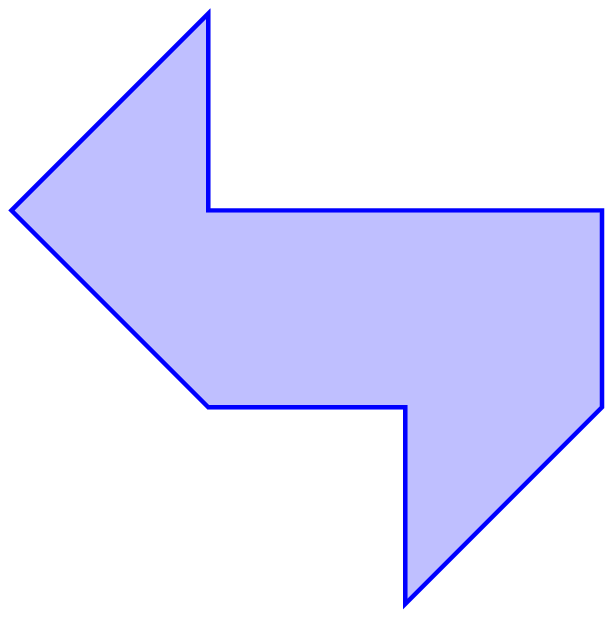}
\hspace{2cm}
\includegraphics[width=4cm]{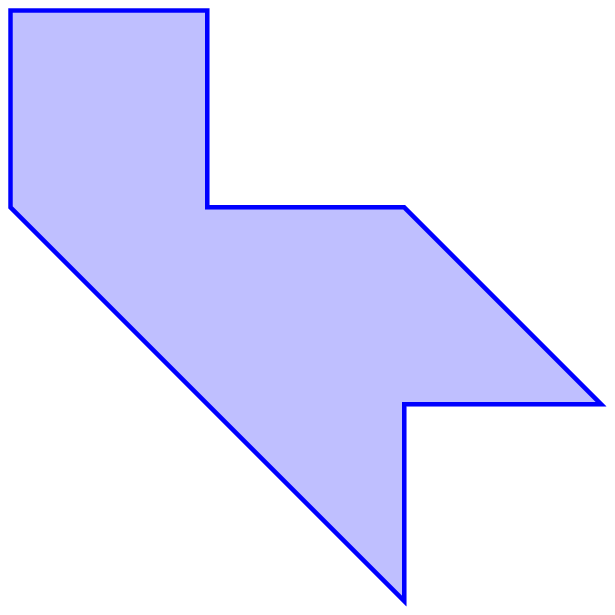}
\caption{Isospectral domains}
\label{Fig_3}
\end{center}
\end{figure}

Consider the domains of Fig.~\ref{Fig_3}. It is known that these domains are isospectral, i.e. that the eigenvalues of the
laplacian on one domain coincide with those on the second domain, as proved by Gordon, Webb and Wolpert~\cite{Gordon92a,Gordon92b}.
The numerical calculation of the eigenvalues of these regions has attracted large interest, using different techniques; for example,
Wu, Sprung and Martorell \cite{Wu95} have used finite difference and mode matching to estimate the first 25 eigenvalues of these
domains; the most precise results have been obtained by Driscoll in Ref.~\cite{Driscoll97}
and by Betcke and Trefethen \cite{Betcke05}. The result that Betcke and Trefethen report for the eigenvalue of the fundamental mode
\begin{eqnarray}
E_1 \approx 2.537943999798 \nonumber
\end{eqnarray}
is slightly more precise than the value reported by Driscoll. Moreover, Sridhar and Kudrolli~\cite{SK94} have performed an experiment
with microwave cavities of the form of the domains of Fig.~\ref{Fig_3}, verifying their isospectrality~\footnote{Readers interested in
the topic of isospectrality should refer to the recent review paper of Giraud and Thas~\cite{Giraud10}.}.

In this case, we have applied finite differences calculating the lowest eigenvalues of both domains for 30 grids; the grid spacing is
chosen appropriately so that the border is sampled exactly~\footnote{With respect to the case of the L-shape, here the domains do not have any
symmetry and only specific grids sample the border; this explains the smaller number of grids which could be used.}. Remarkably, the matrices
obtained with finite difference for the two domains are also isospectral.

\begin{figure}
\begin{center}
\includegraphics[width=8cm]{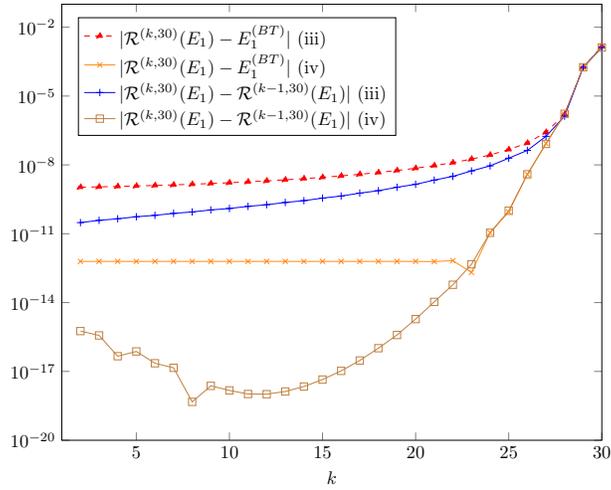}
\caption{Error over the first eigenvalue of the isospectral regions. The first two curves report the difference between the values obtained with Richardson extrapolation of $30-k$ grids, respectively using scheme ${\rm iii}$ and ${\rm iv}$, and the precise value of Betcke and Trefethen \cite{Betcke05}
($E_1 \approx 2.537943999798$); the last two curves report the difference between the values obtained with Richardson extrapolation of $30-k$ grids and the values obtained with Richardson extrapolation of $30-k-1$ grids, respectively using scheme ${\rm iii}$ and ${\rm iv}$. This difference essentially provides the number of stable digits achieved. }
\label{Fig_iso_error}
\end{center}
\end{figure}

In Fig.~\ref{Fig_iso_error} we report the error over the first eigenvalue of the isospectral domains, while in Table \ref{tab_iso} we report
our best estimates for the lowest 25 eigenvalues, obtained using Richardson extrapolation, with the same exponents as for the L.
For the lowest eigenvalue we gain 5 digits with respect to the result of Betcke and Trefethen
\begin{eqnarray}
E_1 = 2.53794399979862045
\end{eqnarray}
Moreover, even our poorest result, for the 25th mode, has two extra digits with respect to the result of Driscoll.

In light of these results, we  stress that the finite difference method can provide very accurate results, despite
the common prejudices. In the abstract of the paper of Driscoll, for example, we read:
"Furthermore, standard numerical methods for computing the eigenvalues, such as adaptive finite elements, are highly inefficient".

A second comment regards the work of Wu, Sprung and Martorell, who calculated the FD eigenvalues for these domains for 3 grids and
then used Richardson extrapolation to obtain better estimates. Incorrectly, they assumed that the FD results vary quadratically with
the grid spacing, a behavior which is appropriate only for the modes of the square (modes 9 and 21).

\begin{table}[!htbp]
\caption{Lowest 25 eigenvalues of the isospectral domains obtained using Richardson extrapolation with set ${\rm iv}$
(the sets marked with $\dagger$ are eigenstates of the square and are extrapolated using set ${\rm i}$).}
\bigskip
\label{tab_iso}
\begin{center}
\begin{tabular}{|c|l|}
\hline
$n$ & $E_n^{({\rm Richardson})}$ \\
\hline
1            &  2.53794399979862045  \\
2            &  3.65550971352441826  \\
3  &  5.17555935622451540  \\
4            &  6.53755744376443310  \\
5            &  7.2480778625641275588 \\
6            &  9.20929499840321242 \\
7            &  10.59698569133316780 \\
8            &  11.5413953955859566289 \\
9$^\dagger$  &  12.33700550136169827354311374984518891914212 \\
10           &  13.0536540557280658   \\
11           &  14.313862464291008706 \\
12           &  15.871302620009314 \\
13$^\dagger$ &  16.941751687972089 \\
14           &  17.6651184368431201 \\
15           &  18.9810673876525993 \\
16           &  20.882395043282328 \\
17           &  21.2480051773728  \\
18           &  22.23285179297328 \\
19           &  23.711297484824032 \\
20           &  24.479234069273887 \\
21$^\dagger$ &  24.674011002723396547086227499690377838284 \\
22           &  26.08024009965984 \\
23           &  27.304018921125 \\
24           &  28.175128581453 \\
25           &  29.569772913239 \\
\hline
\end{tabular}
\end{center}
\bigskip\bigskip
\end{table}

\subsection{Square domain with a $45^0$-crack}

\begin{figure}
\begin{center}
\includegraphics[width=4cm]{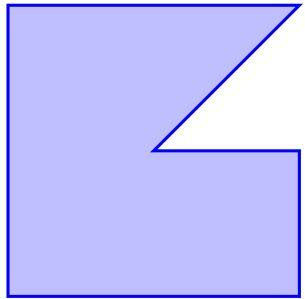}
\caption{Unit square with a $45^0$-crack}
\label{Fig_SlicedSquare}
\end{center}
\end{figure}

The domain represented in Fig.~\ref{Fig_SlicedSquare} is particularly interesting, since it contains a reentrant angle
$\theta = 7\pi/4 $, which is larger than the angle of the L-shaped domain. Additionally, the domain has no symmetry and therefore
the numerical calculation is more demanding than for the case of the L and H shapes.
This problem has been originally studied by Blum and Rannacher \cite{Blum90} and more recently by Yuan and He \cite{He09}, where the bounds
\begin{equation}
35.631515 \leq E_1 \leq 35.631522 \nonumber
\end{equation}
have been obtained. The result $E_1 \approx 35.617$ was obtained in Ref.~\cite{Blum90} applying Richardson extrapolation
to finite elements.

In Table \ref{tab_MPS_sliced} we report the numerical approximations to the lowest 5 eigenvalues of this domain, obtained using
the MPS with 356 points. The digits reported in the table are expected to be correct; in particular for the lowest eigenvalue we
have
\begin{eqnarray}
E_1 \approx 35.63151951719172309520548614207765698409
\end{eqnarray}

\begin{figure}
\begin{center}
\includegraphics[width=4cm]{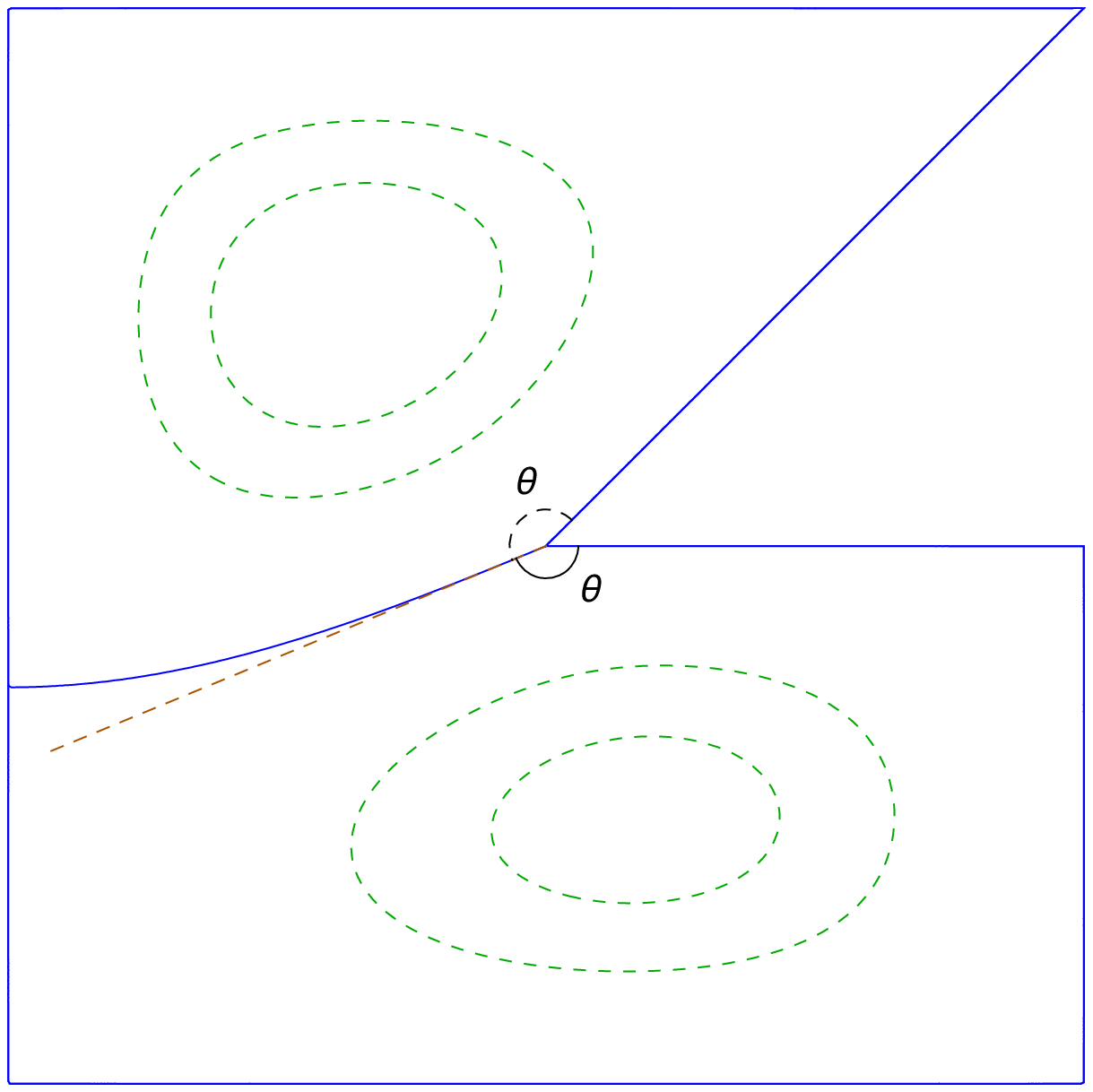} \hspace{1cm}
\includegraphics[width=4cm]{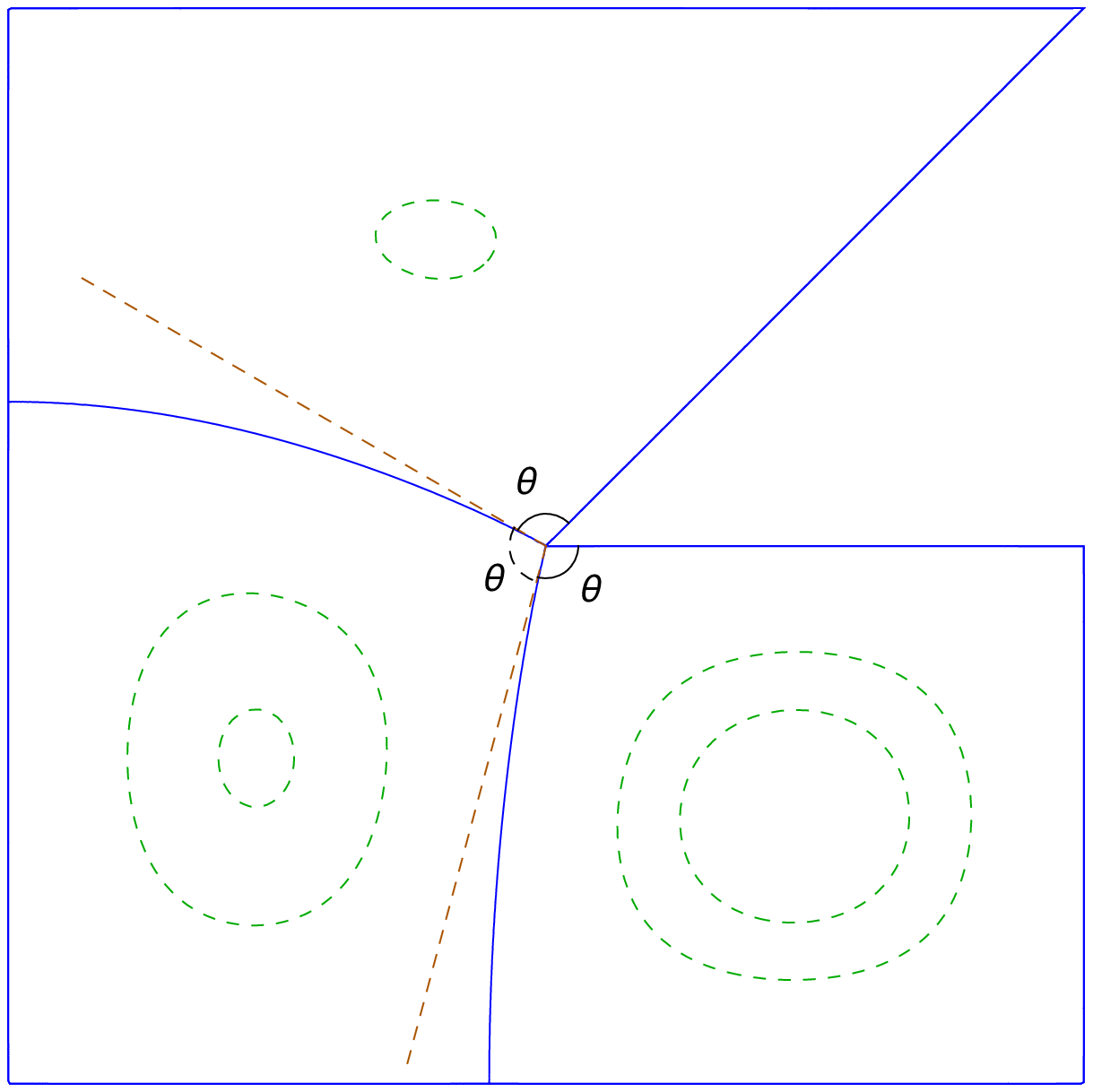} \bigskip \\
\includegraphics[width=4cm]{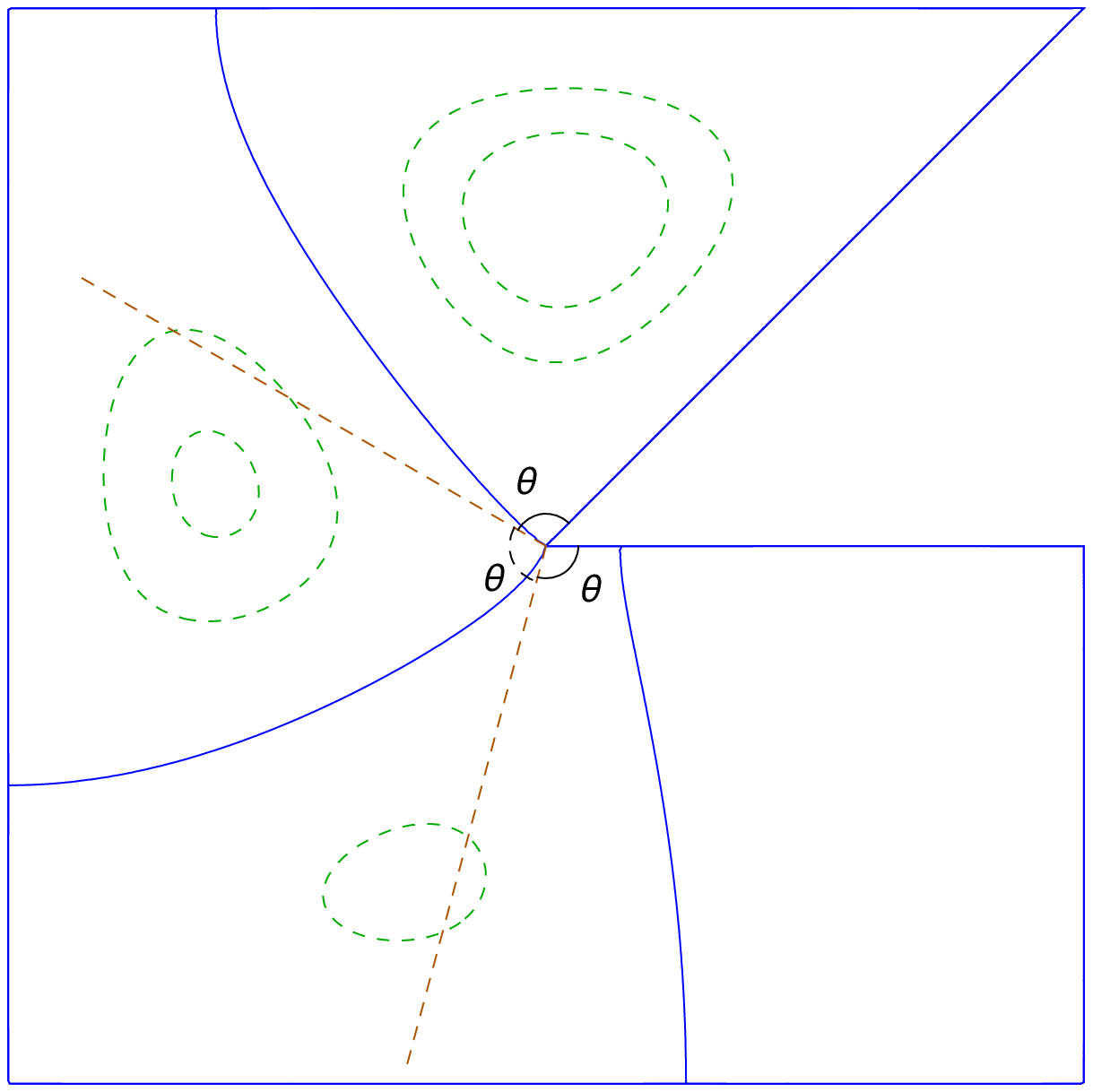} \hspace{1cm}
\includegraphics[width=4cm]{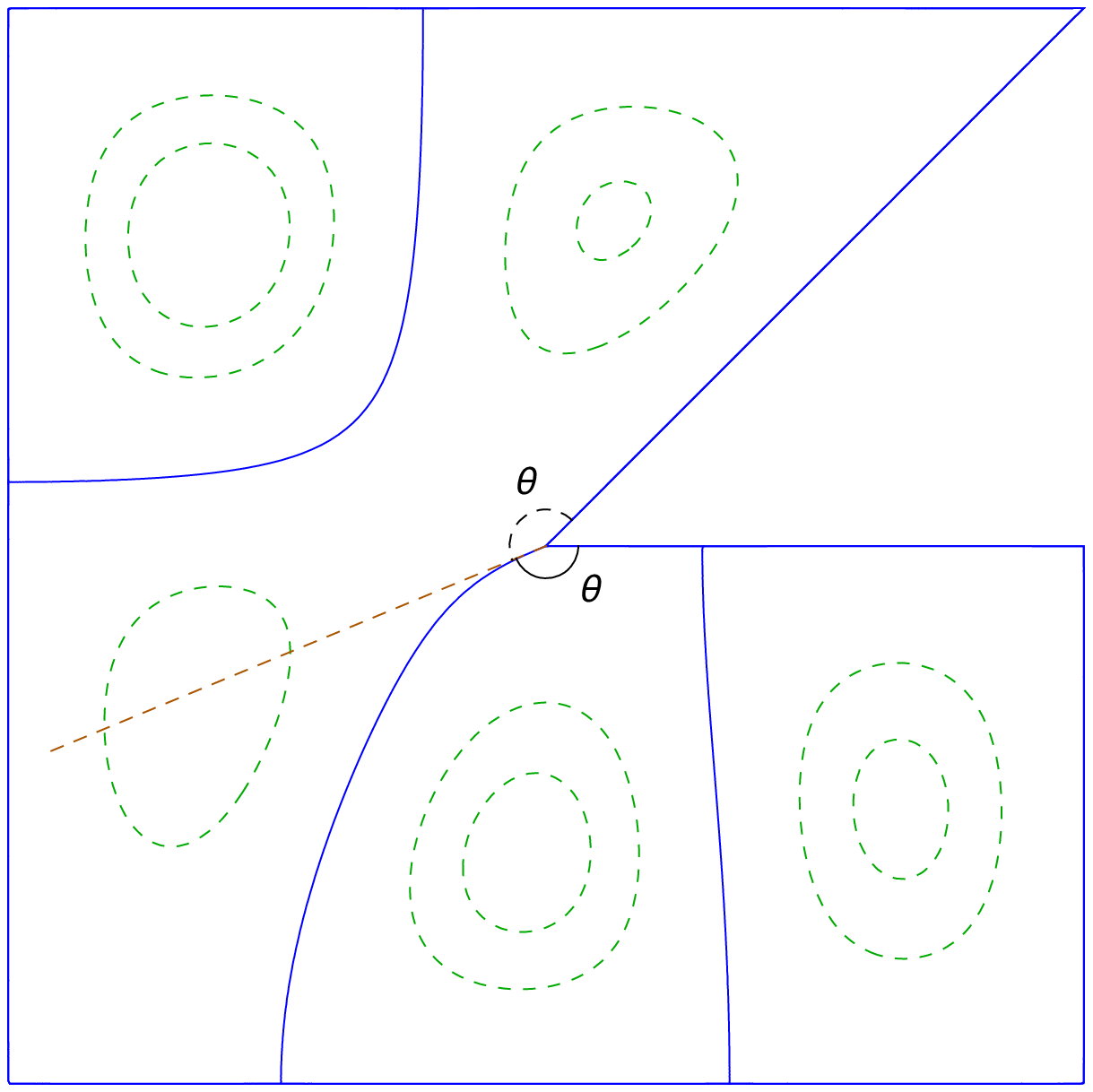}
\caption{Nodal lines of the first four excited modes of the unit square with a $45^0$-crack}
\label{Fig_nodal_SlicedSquare}
\end{center}
\end{figure}

In Fig.~\ref{Fig_nodal_SlicedSquare} we show a contour plot of the first four modes of this domain, obtained using finite differences with
a grid with spacing  $h=1/120$, corresponding to a total of 12331 grid points. The solid blue lines are the nodal lines, while the dashed green lines
are level curves. While the fundamental mode is nodeless, the remaining three states have one or two nodal lines which
start on the vertex of the reentrant corner, thus dividing the original domain in two or more domains. Looking at the figure we see that
for the second state the resulting sub-domains have a reentrant angle $\theta = 7\pi/8$, while for the third and fourth states the sub-domains have a reentrant angle $\theta= 7\pi/12$. The dashed straight lines in the plot are tangent to the nodal line in the vertex.

As a result of this observation, we speculate that the asymptotic behavior of the finite difference eigenvalue may contain the exponents
$8/7$, $16/7$ and $24/7$~\footnote{In the case of the L-shape, the reentrant corner is divided in two halves by the line $y=x$ for the modes
that are odd: in that case, the nodal line is exactly sampled by the grid and therefore the exponent $4/3$ is absent, while the first rational
exponent is $8/3$. In the present case the nodal lines are not sampled by the grid.}.

We have calculated the lowest eigenvalues for this domain using
finite difference with 60 grids; the Richardson and
Richardson-Pad\'e extrapolations of these results, with the
appropriate exponents in the asymptotic series, should allow one
to obtain precise approximations to the eigenvalues of this
domain, as for the case of the L.

\begin{table}[!htbp]
\caption{Lowest 5 eigenvalues of the unit square with a $45^0$-crack obtained with the MPS using $356$ points evenly spaced on the border}
\bigskip
\label{tab_MPS_sliced}
\begin{center}
\begin{tabular}{|c|c|}
\hline
$n$ & $E_n^{({\rm MPS})}$ \\
\hline
1           &  35.63151951719172309520548614207765698409 \\
2           &  54.19310844424629197411978585647040768914 \\
3           &  73.63330812560383459483828674566950026083 \\
4           &  104.3280904734882128897772035674716112638\\
5           &  124.5914636064409738708659060017320376707 \\
\hline
\end{tabular}
\end{center}
\bigskip\bigskip
\end{table}

In this case we have extrapolated the finite difference results using a series of the form
\beq
E(h) &=& E(0)  +  c_1 \ h^{8/7}+c_2 \ h^2 + c_3 \ h^{16/7}+ c_4 \ h^{22/7} + c_5 \ h^{24/7} + c_6 \ h^4 \nonumber \\
&+&  c_7 \  h^{30/7} + c_8 \ h^{32/7} + c_9 \  h^{36/7} + c_{10} \  h^{38/7} +  c_{10} \  h^{40/7} + c_{11} \ h^{6} + c_{12} \ h^{48/7} \nonumber \\
&+& c_{13} \ h^{8} + c_{14} \ h^{64/7} + c_{10} \ h^{72/7}+ c_{16}  \ h^{80/7} + c_{17}  \ h^{12} + c_{18}  \  h^{88/7} \nonumber \\
&+& c_{19}  \  h^{96/7}+ c_{20}  \ h^{104/7}+ c_{21}  \ h^{120/7} + c_{22} h^{128/7} + c_{23}  \  h^{136/7} + c_{24}  \ h^{20/7}\nonumber \\
&+& c_{25}  \  h^{144/7} + c_{26}  \ h^{152/7}+ \dots
\label{asym_squaretriangle}
\eeq
where the coefficients are chosen {\sl empirically} and include the ones mentioned earlier.

\begin{figure}[!htbp]
\begin{center}
\includegraphics[width=8cm]{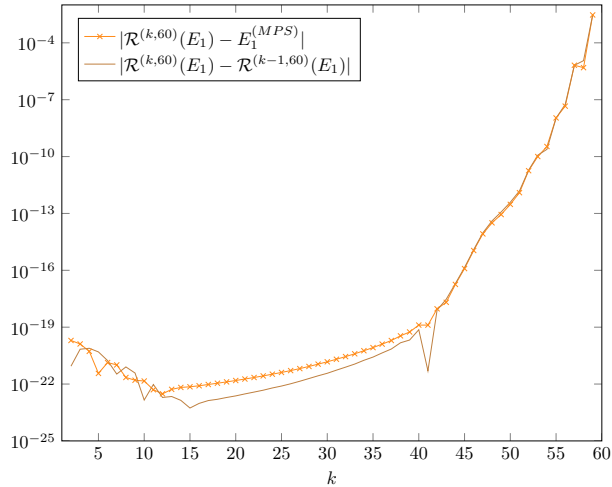}
\caption{Error over the first eigenvalue of the unit square with a $45^0$-crack. The asymptotic series of Eq.~(\ref{asym_squaretriangle})
has been used.}
\label{Fig_sliced_error}
\end{center}
\end{figure}

\begin{table}[!htbp]
\caption{Correct digits of the first 5 eigenvalues of the unit
square with a $45^0$-crack, obtained by  applying the Richardson
and Richardson-Pad\'e extrapolations to FD eigenvalues. }
\bigskip
\label{table_crack}
\begin{center}
\begin{tabular}{|l|c|c|c|c|}
\hline
$n$ &  $\log_{10} \frac{1}{|E_n^{({\rm RE})}-E_n^{\rm (MPS)}|}$ & $\log_{10} \frac{1}{|E_n^{({\rm PRE})}-E_n^{\rm (MPS)}|}$ \\
\hline
1           & 22.18 & 25.37  \\
2           & 23.65 & 23.87  \\
3           & 22.00 & 23.92  \\
4           & 21.04 & 24.22  \\
5           & 20.85 & 23.01  \\
\hline
\end{tabular}
\end{center}
\bigskip\bigskip
\end{table}

It is interesting to check the numerical values obtained for the coefficients of the series (\ref{asym_squaretriangle}), using
the Richardson extrapolation of the FD results corresponding to the last 30 grids, for the modes above:
\beq
E_1(h)  &\approx&  35.63151952  +  22.47641559 \ h^{8/7}-71.03523727 \ h^2 + 6.078713368 \ h^{16/7}\nonumber \\
&-& 78.46323288 \ h^{22/7} -8.840565052 \ h^{24/7}  + 63.35756993 \ h^4 + \dots \\
E_2(h) &\approx& 54.19310844 - 2.87 \times 10^{-17} \ h^{8/7}-164.3992546\ h^2 -21.20457267\ h^{16/7} \nonumber \\
&+& 1.03 \times 10^{-8} \ h^{22/7}  -1.44 \times 10^{-7} \ h^{24/7} + 212.7295338\ h^4 + \dots \\
E_3(h) &\approx& 73.63330813  +  3.52 \times 10^{-17} \ h^{8/7}- 260.5413126 \ h^2 +  8.56 \times 10^{-12} \ h^{16/7}\nonumber \\
&-& 3.15 \times 10^{-8} \ h^{22/7} -91.25393089\ h^{24/7} + 222.794824 \ h^4 + \dots \\
E_4(h) &\approx& 104.3280905   -2.12 \times 10^{-15} \ h^{8/7}  -668.8593013 \ h^2   -3.38 \times 10^{-10} \ h^{16/7} \nonumber \\
&+& 9.07 \times 10^{-7} \ h^{22/7}  -39.10703889 \ h^{24/7} + 1997.967306 \ h^4 + \dots \\
E_5(h) &\approx& 124.5914636  -2.6 \times 10^{-15} \ h^{8/7} -766.4031071 \ h^2 -13.2187842 \ h^{16/7} \nonumber \\
&+& 1.17 \times 10^{-6} \ h^{22/7}
-0.00001758167793 \ h^{24/7} + 1901.063425 \ h^4 + \dots
\label{asym2b}
\eeq

Clearly one observes that depending on the mode chosen, some of the coefficients are consistent with a vanishing value: these observations
are summarized in Table \ref{table_crack_2}, where the leading rational coefficients and the corresponding reentrant angle
are reported for each of the first 5 modes.

\begin{table}[!htbp]
\caption{Leading rational exponents of the FD series for the first 5 modes of the square with a $45^0$-crack, and corresponding
reentrant angles. }
\bigskip
\label{table_crack_2}
\begin{center}
\begin{tabular}{|l|c|c|c|c|}
\hline
$n$ &  leading exponent & dominant angle \\
\hline
1           & $\frac{8}{7}$ & $\frac{7\pi}{4}$ \\
2           & $\frac{16}{7}$ & $\frac{7\pi}{8}$ \\
3           & $\frac{24}{7}$ & $\frac{7\pi}{12}$ \\
4           & $\frac{24}{7}$ & $\frac{7\pi}{12}$ \\
5           & $\frac{16}{7}$ & $\frac{7\pi}{8}$ \\
\hline
\end{tabular}
\end{center}
\bigskip\bigskip
\end{table}

\subsection{Square domain with two slits}

\begin{figure}[!htbp]
\begin{center}
\includegraphics[width=4cm]{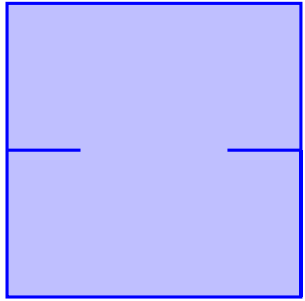}
\caption{Square domain with two slits}
\label{Fig_Slit}
\end{center}
\end{figure}

Consider the unit square with two $1/4$ slits, represented in Fig.~\ref{Fig_Slit}.
This example has been studied in Refs.~\cite{Blum90, Liu09}.
In this case the re-entrant corner is $2 \pi$, thus the leading exponent in the FD series
is $\alpha_1= 1$. Eliminating the pollution of this contribution, Blum and Rannacher were
able to obtain $E_1 = 35.728$ for their finest grid.

Consistently  with our previous assumptions, we conjecture that the FD series has the form
\begin{eqnarray}
E^{(k)} = c_0 + \sum_{j=1}^\infty c_j h_k^{j}
\label{series_FD_split}
\end{eqnarray}
which is the typical form used in Richardson extrapolation. In this case, Bender and Orszag provide in
\cite{bender1999advanced} a nice explicit formula for the coefficient $c_0$ (Eq.(8.1.16) of pag. 375 of their book), which in our notation reads:
\begin{eqnarray}
c_0 = \sum_{k=0}^N \frac{E^{(n+k)} (n+k)^N (-1)^{k+N} }{k! (N-k)!}
\end{eqnarray}

Our numerical experiments with this domain consist of two sets:

\begin{itemize}
\item a set which contains the numerical approximation to the lowest eigenvalue of the domain calculated to
220 digits of accuracy using the CGM, for 36 grids with $h = 1/2n$ and $n=8, 10, \dots, 80$;

\item a set which contains the numerical approximation to the lowest 50 eigenvalues of the domain calculated
to 60 digits arithmetic using the internal Mathematica command \verb Eigenvalue    for 20 grids with
$h = 1/2n$ and $n=8, 10, \dots, 46$;
\end{itemize}

In table \ref{tab_slit} we report the approximate values of selected eigenvalues of this domain, obtained using
Richardson and Pad\'e-Richardson extrapolation. The eigenvalue of the fundamental mode is obtained using the
first set of FD results, whereas the remaining eigenvalues are obtained using the second set.
The digits reported in the table are believed to be correct. The table omits the  eigenmodes
of the square, for which the convergence is much faster.

\begin{table}[!htbp]
\caption{Selected eigenvalues of the square with two slits obtained using Richardson and Pad\'e-Richardson extrapolation
of the FD results}
\bigskip
\label{tab_slit}
\begin{center}
\begin{tabular}{|c|l|l|}
\hline
$n$ & $E_n^{({\rm R})}$ & $E_n^{({\rm PR})}$ \\
\hline
1  &  28.131367480845754755206 & 28.131367480845754755206268 \\
3  &  70.65038470368   & 70.65038470368488 \\
5  &  99.846759253895  & 99.8467592538950  \\
7  &  130.483305932580 & 130.4833059325804 \\
8  &  153.39663535893  & 153.3966353589373 \\
10 &  196.598428600514 & 196.5984286005142 \\
13 &  218.04116455831  & 218.0411645583168 \\
15 &  268.2038796851519& 268.2038796851519 \\
16 &  272.5993876495   & 272.59938764953   \\
17 &  280.750584654    & 280.7505846542989 \\
20 &  348.460286264284 & 348.4602862642840 \\
50 &  750.8475130      & 750.847513086     \\
\hline
\end{tabular}
\end{center}
\bigskip\bigskip
\end{table}

\begin{figure}
\begin{center}
\includegraphics[width=5cm]{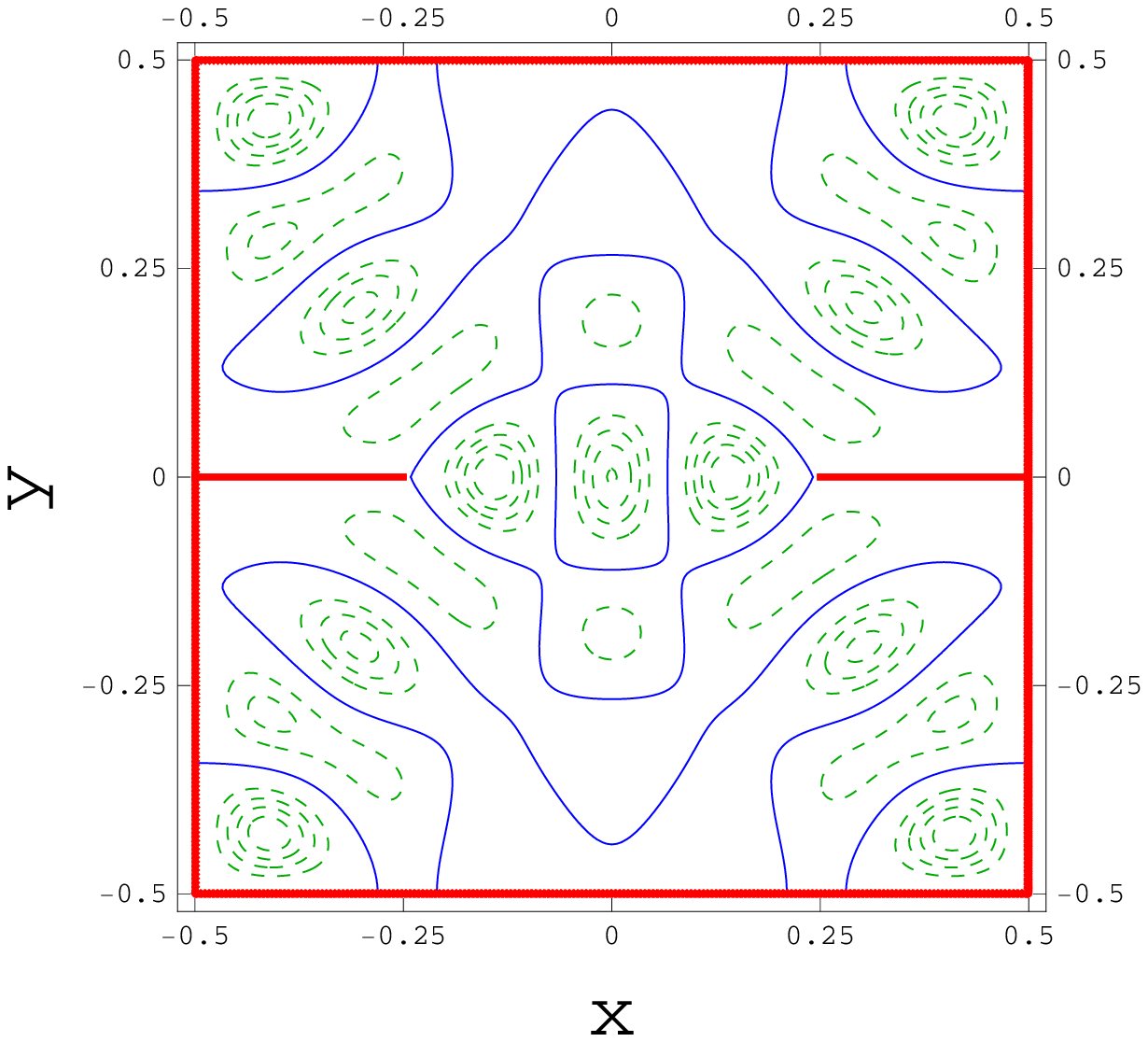}
\hspace{1cm}
\includegraphics[width=5cm]{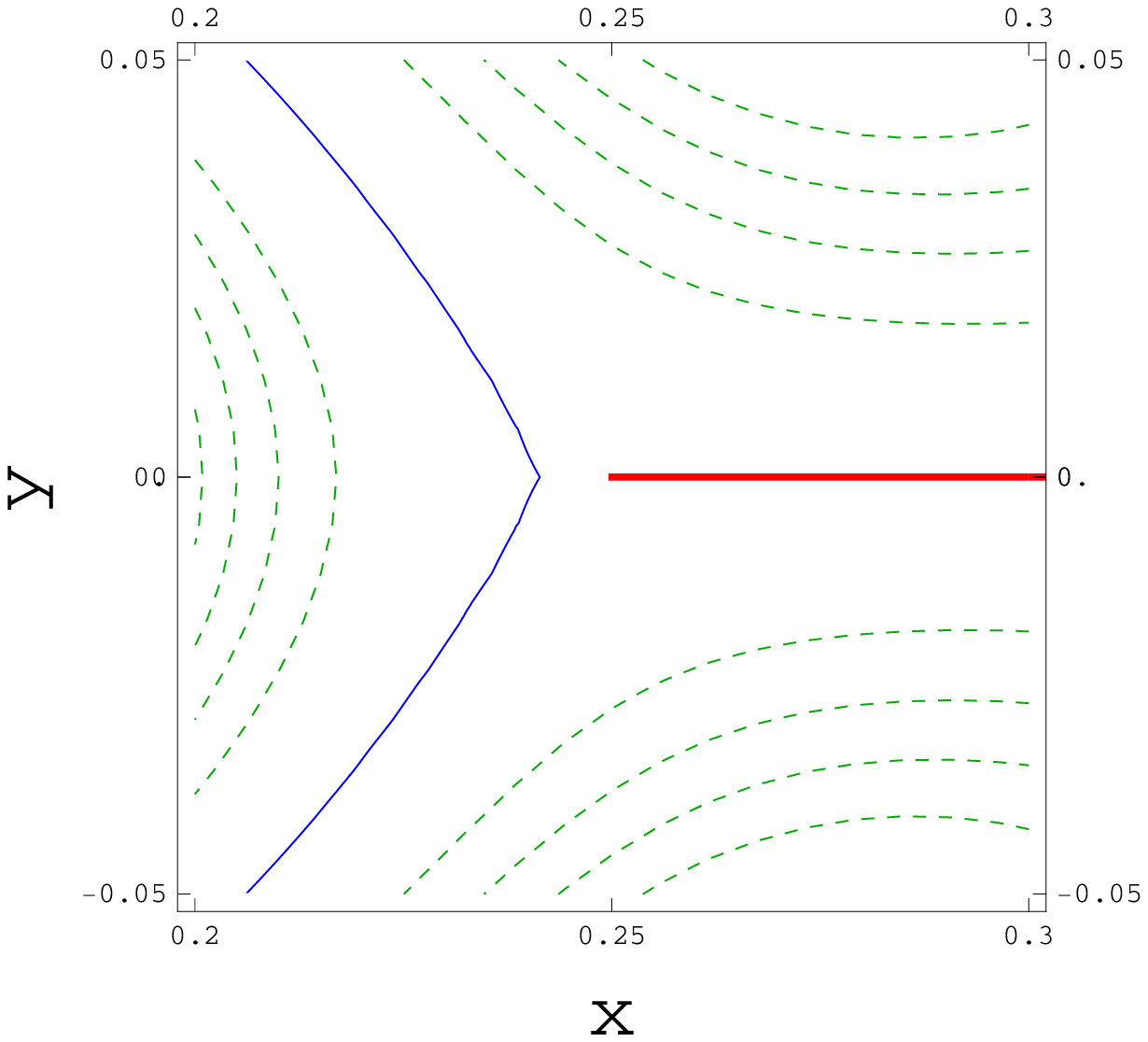}
\caption{Nodal lines of the $50^{\rm th}$ mode of the square domain with two slits.}
\label{Fig_slit50}
\end{center}
\end{figure}

Of particular interest is the fiftieth mode, whose nodal lines are the solid lines displayed in Figs.~\ref{Fig_slit50}.
Looking at the left plot, we are tempted to assume that a nodal line partitions each of the $2\pi$  reentrant angles into
three angles of $2\pi/3$, which would imply that the corresponding FD series would now have rational exponents.
A simple analysis of the FD results however shows that this mode is also described by the series in eq.~(\ref{series_FD_split}).
This behavior is consistent with the information delivered by the right plot in Figs.~\ref{Fig_slit50}, that reveals that in fact
the nodal line {\sl do not end} in the reentrant corner. In other words, the study of the FD series for a given domain, can also
provide information on the behavior of the nodal lines of the corresponding eigenmodes.

\section{Conclusions}
\label{conclusions}

In this paper we have showed that it is possible to obtain precise estimates for the eigenvalues of the negative
Laplacian over particular domains in the plane  by  performing a Richardson extrapolation or a rational (Pad\'e)-Richardson
extrapolation of the results obtained with finite differences, where the exponents of the series are related to the
reentrant angles in the domain.
The problem of determining the series describing the behavior of the finite difference
results from first principles is difficult and it seems that a theoretical study is still lacking. The problem
is both challenging and interesting for the  applications of finite differences in  Physics,
Applied Mathematics and Engineering are as numerous as the stars in the Milky Way. Quoting Kuttler and Sigillito, pag. 178 of \cite{Kuttler84},
"the exact form of the first several terms in the asymptotic formula for specific regions where no boundary interpolation
is required is a nice problem at about the level of a doctoral thesis." The fact that, since 1984 this problem has not been
yet solved suggests an even higher level of difficulty.

In this paper we have pursued the less ambitious goal of identifying the series (i.e. the exponents) empirically and we have
obtained particularly encouraging results. In the case of the L-shaped domain, for instance, the extrapolation of the results
obtained with finite differences leads to a determination of the first 68 digits of the lowest eigenvalue.

The knowledge of the finite difference series for a given domain allows a precise determination of the numerical values
of the eigenvalues of that domain, making the finite difference method a powerful computational
tool~\footnote{In all the examples that we have treated in this paper, we have been able to improve published results.}.

Here we stress the most relevant observations obtained from a careful analysis of the numerical results for the examples
considered in this paper:
\begin{itemize}
\item The FD series appears to be an asymptotic series, as suggested by the particular behavior of the error; this does not
limit the accuracy of the extrapolated results, if the largest spacing of the set is appropriately decreased, as more and more
terms are added;
\item The example of the square with a $45^0$ crack tells us that when a nodal line terminates in a reentrant corner,  the corresponding
FD series have exponents corresponding to the fractions of reentrant angles, even if the nodal line is not completely sampled by the grid
(it is the behavior infinitesimally close to the corner that matters);
\item It is reasonable to assume that, for a given domain, the FD series corresponding to the different modes all are described by the same
series (although for some modes some exponents could be missing for symmetry reasons -- this is the case of the modes
of the L which are also eigenmodes of the square, for which all the coefficients of all rational exponents vanish );
\item If the observation above is correct, this means that one cannot have nodal lines partitioning the reentrant corner
if the new exponent generated is not of the type already contained in the series! The case of the fiftieth mode of the square with two slits
illustrates this behavior: the nodal lines stretch almost completely to the reentrant corner, although they do not join it!
\item We conjecture that the nature of the reentrant corners fully determines the exponents of the FD series and therefore different
domains, containing the same reentrant angles should all have the same exponents (see for example the case of the L, of the H and
of the isospectral domains considered in this paper); this makes Richardson (and Richardson-Pad\'e) extrapolation practical
even for complicated domains  where the use of MPS can be problematic;
\item For the case of the L-shape and of the square with a $45^0$ crack,  our results also provide an independent check/validation of the
corresponding results obtained using MPS;
\end{itemize}

\section*{Acknowledgements}
The research of P.A. was supported by Sistema Nacional de Investigadores (M\'exico).

\bibliographystyle{siam}  

\bibliography{Richardson_Extrapolationi_bibtex_referrences}  

\end{document}